%% file: EM_PIC_paper.tex
\documentclass[12pt]{iopart}
\input{customed_settings}
\usepackage[dvipsnames]{xcolor}
\usepackage{todonotes}
\usepackage{cite}

\begin{document}

%\title[Modeling of VHF CCP with a fully EM PIC code]{Modeling of very high frequency capacitively coupled plasmas with a fully electromagnetic particle-in-cell code}

\title[Modeling of VHF large-electrode CCPs with a fully electromagnetic PIC]{Modeling of very high frequency large-electrode capacitively coupled plasmas with a fully electromagnetic particle-in-cell code}  

\author{D. Eremin$^{1}$, E. Kemaneci$^{1}$, M. Matsukuma$^{2}$, T. Mussenbrock$^{3}$, R.P. Brinkmann$^{1}$}
\address{$^1$ Institute of Theoretical Electrical Engineering, Ruhr University Bochum, Universitätsstrasse 150, D-44801 Bochum, Germany}
\address{$^2$ Tokyo Electron Technology Solutions Limited, 2381-1 Kitagejo, Fujii-cho, Nirasaki City, Yamanashi 407-8511, Japan}
\address{$^3$ Chair of Applied Electrodynamics and Plasma Technology, Ruhr University Bochum, Universitätsstrasse 150, D-44801 Bochum, Germany}

\ead{denis.eremin@rub.de}
\vspace{10pt}
\begin{indented}
\item[]
\today
\end{indented}

\begin{abstract}
Phenomena taking place in capacitively coupled plasmas with large electrodes and driven at very high frequencies
are studied numerically utilizing a novel energy- and charge-conserving implicit fully electromagnetic particle-in-cell / Monte Carlo code ECCOPIC2M. 
The code is verified with three model problems and is validated with results obtained in an earlier experimental work \cite{sawada_2014}. 
The code shows a good agreement with the experimental data in four cases with various collisionality and
absorbed power. It is demonstrated that under the considered parameters, the discharge produces radially uniform
ion energy distribution functions for the ions hitting both electrodes. In contrast, ion fluxes exhibit a strong radial
nonuniformity, which, however, can be different at the powered and grounded electrodes at increased pressure. It
is found that this nonuniformity stems from the nonuniformity of the ionization source, which is in turn shaped
by mechanisms leading to the generation of energetic electrons. The mechanisms are caused by the interaction of electrons
with the surface waves of two axial electric field symmetry types with respect to the reactor midplane. The asymmetric modes
dominate electron heating in the radial direction and produce energetic electrons via the relatively inefficient
Ohmic heating mechanism. In the axial direction, the electron energization occurs mainly through an
efficient collisionless mechanism caused by the interaction of electrons in the vicinity of an expanding sheath with the sheath motion,
which is affected by the excitation of the surface modes of both types. The generation of energetic electron populations
as a result of such mechanisms is shown directly. Although some aspects of the underlying physics were demonstrated in the
previous literature with other models, the particle-in-cell method is advantageous for the predictive modeling
due to a complex interplay between the surface mode excitations and the nonlocal physics of the corresponding type of
plasma discharges operated at low pressures, which is hard to reproduce in other models realistically.
\end{abstract}

%
% Uncomment for keywords
\vspace{2pc}
%
% Uncomment for Submitted to journal title message

%\submitto{\PSST}

%
% Uncomment if a separate title page is required
%\maketitle
% 
% For two-column output uncomment the next line and choose [10pt] rather than [12pt] in the \documentclass declaration
%\ioptwocol
%

% \textcolor{purple}{Thomas' comments}
% \newcommand{\commFP}[1]{\textcolor{blue}{#1}}
% \commFP{Federico's comments}

%%%%%%%%%%%%%%%%%%%%%%
%\clearpage
\section{Introduction} \label{sec1}

Capacitively coupled plasma discharges represent one of the main tools currently used in plasma processing, especially in 
the dry etching and plasma enhanced chemical vapor deposition technologies \cite{lieberman_2005,chabert_2011,makabe_2006}.  
There is a continuing trend towards increasing the size of a processed wafer and towards including a very high frequency ($\geq 30$ MHz) harmonic
in the driving voltage to increase the ion flux impinging on the wafer while keeping the ion energy low. This can be explained as follows: Since the power
absorbed by the plasma is approximately quadratically proportional to the driving frequency and only linearly proportional to
the voltage amplitude, by utilizing a higher frequency it is possible to obtain approximately the same plasma density and 
the ion flux at a substantially reduced voltage. At low neutral gas pressures typical for the mentioned technologies, ion energies at the electrodes are determined by the sheath voltage drop being proportional to the driving voltage amplitude.

Unfortunately, the increase in the driving frequency and the electrode size leads to nonuniformities of various discharge properties
relevant to the plasma processing technologies (e.g., \cite{stevens_1996,schmitt_2002,perret_2003}).
This has been attributed to the increase of the ratio between the electrode size and the 
wavelength of the surface modes inherent to a plasma sandwiched by the electron-depleted sheaths (Strictly speaking, the corresponding modes have the appearance of truly ``surface'' waves in the sense of having electromagnetic fields concentrated close to the sheath boundaries only if the skin depth is much smaller than the electrode distance). Since such a wavelength is much smaller compared
to the vacuum wavelength at the same frequency, it can become comparable to the reactor size at much smaller frequencies. 
Two different fundamental types of the surface modes sustained by a plasma-filled CCP reactor were found, both propagating along the electrodes. 
In a geometrically symmetric reactor, the modes can be classified by their axial
electric field symmetry with respect to the reactor midplane. The ``symmetric'' or ``even'' wave has the axial electric field aligned
\cite{bowers_2001,lieberman_2002,sansonnens_2006,kawamura_2018,dvinin_2021}, and the axial electric field in the ``antisymmetric'' or ``odd'' wave
is opposed \cite{bowers_2001,howling_2004,sansonnens_2006,kawamura_2018,dvinin_2021} in the upper and the lower reactor halves. 
For each of the fundamental modes, there is a set of evanescent modes which can be non-negligible at the edge, but are strongly damped in 
the direction of the radial center \cite{lieberman_2002,sansonnens_2006}. These modes are needed to satisfy boundary conditions for the
electromagnetic fields at the reactor edge \cite{lieberman_2002} in case the fundamental mode does not behave appropriately there.
The sheath voltage radial profile far from the electrode edge is determined by the fundamental symmetric and asymmetric modes
triggered at the excitation frequency, and their radial wavenumbers can be obtained by calculating the corresponding dispersion relations at
the excitation frequencies. 
%At relatively low frequencies below $30$ MHz the symmetric mode typically dominates the excitation. 

The analytic treatment was later generalized from the symmetric case with equal sheath width and equal electrode areas to a more
realistic geometry with the powered electrode having a smaller area and a larger sheath width in \cite{lieberman_2016} (see also \cite{dvinin_2021}).
Although more complicated geometrical layouts require the introduction of additional parameters for obtaining consistent field patterns, the
patterns remain the same concerning the field topology and thus can still be easily distinguished.
From the physics standpoint, these modes are also responsible for different fundamental phenomena in the CCP discharges. The symmetric
mode is related to the plasma series resonance, and the antisymmetric mode affects the self-bias \cite{howling_2004,howling_2005}, which could also serve as a criterion for
their naming \cite{eremin_2017b}. Based on the established terminology,
the symmetric mode is almost always meant in the context of the ``standing wave effect'' 
in CCP discharges (e.g., \cite{lieberman_2002}), although, strictly speaking, both the symmetric and the antisymmetric modes can form standing waves \cite{howling_2007}, see also Fig.~\ref{Fig_mode_identification_sym_and_asym_voltages} later in this paper. The standing wave formed
by the symmetric mode leads to a radial dependence of the total voltage across the electrodes and has a characteristic center-peaked profile. The
antisymmetric mode is sometimes mentioned in relation to the ``telegraph effect'' (e.g.,\cite{howling_2004,howling_2005}), which seems to go back
to the fact that the corresponding mode was first described with the help of the telegraph equation governing signal propagation along
a transmission line.  This, however, is also somewhat ambiguous since the symmetric mode can be described by a telegraph-like
equation for a transmission line as well (e.g.,\cite{lieberman_2002,chabert_2004}). For long wavelengths, the symmetric mode dispersion relation is significantly different 
in the electrostatic (ES) approximation, where it has a cut-off at the plasma series resonance (PSR) frequency \cite{bowers_2001,eremin_2017b} (in fact, the electrostatic dispersion relation for the symmetric mode also includes a vertical section $0 < \omega <  \omega_{pe}\sqrt{s/l} \, , \, k = 0$)
and the fully electromagnetic (EM) treatment, where it has a ``knee'' and goes to zero frequency in the limit of infinitely large wavelengths. 
In contrast, unless the plasma densities are so large that the skin effect comes into play, the antisymmetric mode dispersion relation is
virtually the same in the electrostatic and electromagnetic descriptions  \cite{bowers_2001,eremin_2016,eremin_2017a,eremin_2017b,wen_2017b}. Early single-harmonic treatments predicted that for a sufficiently large plasma density (so that the skin effect starts to play a significant role), the Ohmic heating caused by the 
radial electric field should dominate over the axial electric field Ohmic heating, and the plasma density profile should change from the center-peaked
to the edge-peaked \cite{lieberman_2002,chabert_2006}. This, however, contradicted experimental observations that the profile becomes
more sharply center-peaked with growing power \cite{upadhyay_2013,sawada_2014}. The suggested
explanation attributed this phenomenon to the emergence of higher harmonics of the fundamental driving frequency detected in the plasma
bulk \cite{miller_2006,upadhyay_2013,sawada_2014,lane_2016,zhao_2019,zhao_2021,liu_2021a}. Because the higher harmonics have smaller surface mode wavelengths, they should
lead to enhanced plasma nonuniformities even when a quarter of the surface mode wavelength at the fundamental frequency is larger than
the electrode size. It was also observed in \cite{sawada_2014} that the higher harmonics were present predominantly in the central region, whereas
spectrum of the detected electromagnetic activity at the discharge periphery rapidly fell, which should also lead to a radially
nonuniform plasma density profile having a peak at the center.

A later theoretical work \cite{lieberman_2015} linked the excitation of higher harmonics observed in the aforementioned experimental works to the nonlinearities in the sheath motion
analogous to the PSR phenomenon previously observed experimentally in, e.g., \cite{annaratone_1995,klick_1996}, and typically considered in the electrostatic approximation under the assumptions of
the plasma density and the sheath voltage radial uniformity \cite{Ku_1998a,Ku_1998b,mussenbrock_2006,czarnetzki_2006} (The work \cite{mussenbrock_2007}, which treated the PSR analytically using the full electromagnetic description, did not explicitly address the issues discussed here). Initially, it was found that nonlinearities in the sheath motion of a strongly asymmetric reactor led to a sudden change in the total current when the sheath starts to expand after the complete collapse. This was believed to generate a broad spectrum of the total current harmonics, which caused self-excitation of the PSR oscillations related to the resonance between the capacitive sheath and the inductive plasma bulk. Later it was discovered that the PSR oscillations can be also triggered in geometrically symmetric CCPs due to the cubic nonlinearities in the sheath charge-voltage characteristics and the time dependence of bulk plasma frequency \cite{schuengel_2015}.  
The paper \cite{lieberman_2015} distinguishes between the PSR phenomenon and the excitation of the symmetric mode. However, as mentioned above, it is the same resonance leading to the emergence of the symmetric mode's cut-off frequency 
in the electrostatic dispersion relation at $k=0$, meaning the radial uniformity. Adopting this view that the PSR phenomenon is in fact a result of the symmetric mode's excitation triggered by the broad spectrum of the total current harmonics generated at the moment of the sheath collapse, one can argue that the corresponding
phenomenon should be somewhat modified in the electromagnetic treatment since the symmetric mode dispersion relation is different in the electrostatic
and the fully electromagnetic descriptions. In the latter case,
the dispersion curve $\omega(k)$ for this mode remains close to the PSR frequency until a relatively large wavelength,
%[IT WOULD BE NICE TO ESTIMATE THE K WHERE THE KNEE OCCURS]
whereupon the curve goes to zero, forming the knee \cite{bowers_2001}. 
The wavenumbers of the excited surface modes governing the uniformity of discharge properties are determined by the EM dispersion relations $k(\omega)$ for the surface modes of both symmetries at the driving frequency harmonics contained in the excitation spectrum. It can also be expected that the plasma-filled reactor should exhibit resonant power absorption when the radial wavenumber thus obtained
leads to fulfillment of the boundary conditions for these modes \cite{lieberman_2015,lieberman_2016} (see also \cite{eremin_2017a,eremin_2017b}). 
Otherwise, the boundary conditions are satisfied with the help of the evanescent modes \cite{lieberman_2002} (note that it is virtually always the case in the electrostatic approximation for the symmetric mode excited at the driving frequency). An interesting related result is that by resonantly driving the fundamental asymmetric mode, it is possible to sustain a CCP discharge with asymmetric properties in a symmetric reactor \cite{kawamura_2018}.

Many different approaches have been utilized for the numerical modeling of electromagnetic phenomena in VHF CCP discharges. Some of the works
aimed at expanding the analytical model based on a linear transmission model for the fields in the frequency domain, 
accounting for the plasma electromagnetic properties by its dielectric permittivity function. They introduced a more consistent plasma
model coupled to the field model \cite{chabert_2004,chabert_2006,chabert_2007,lee_2008,kawamura_2011,kawamura_2014}. The advantage of such an approach is that it can reproduce the correct behavior of both mode dispersion relations since the plasma dielectric permittivity takes into account the electron inertia,
which is typically problematic in the fluid-based models. Among the major limitations are the lack of nonlinear sheath dynamics, which drives the modes at harmonics of the fundamental frequency, and the absence of nonlocal and kinetic effects, which could be substantial at low pressures. Other self-consistent models were based 
on the fluid description of plasma and treated the electromagnetic fields in the time domain (e.g., \cite{rauf_2008,bera_2009,yang_2010,zhang_2010,upadhyay_2013}),
which could potentially allow for some nonlinear effects. However, such models typically describe the electron dynamics using the drift-diffusion approximation,
which neglects the electron inertia effects. As it was argued in \cite{eremin_2017a}, such techniques have problems capturing the Ohmic power absorption for the
symmetric mode and the dispersion relation for the asymmetric mode. Although a fluid model was build using the Darwin approximation for the description of electromagnetic fields and taking into account the electron inertia \cite{eremin_2013}, it was realized that it has a limited applicability due to nonlocal electron transport at low pressures and the fact that the Darwin approximation might become inapplicable if sufficiently high harmonics of the fundamental frequency are excited \cite{upadhyay_2013}.

%, the latter having strong impact of the plasma dynamics \cite{lieberman_2016,kawamura_2018}. 
Without the electron inertia effect, the nonlinear sheath dynamics causing the generation of excitation frequency harmonics
cannot be correctly modeled. A more recent class of models has addressed this issue by coupling the transmission line model to a description of
nonlinear sheath dynamics in the time domain \cite{lieberman_2015,wen_2017,wen_2017a,zhao_2019,liu_2021,liu_2021a,liu_2022}. Such models are capable of reproducing the generation of higher harmonics of the fundamental excitation frequency, but miss potentially significant effects inherent to low pressure
CCP plasmas, such as the nonlocal transport of energetic electrons produced, for example, due to the sheath motion or secondary electron emission. 
A recent work \cite{rauf_2020} is one of the examples emphasizing the importance of such effects for low-pressure plasmas, making a fluid 
description inadequate.

%According to the particle continuity equation, the plasma density profile forms as a result of two counteracting processes - production of new particles through 
%the impact ionization and particle transport. As the electric field in the bulk is small, the dominant transport process there is the ambipolar diffusion. 
%At low pressures, no matter what the power deposition profile, the non-local diffusion leads to the center-peaked profile of the electron density (e.g., \cite{chabert_2007,rauf_2020}). 
 
One could therefore argue that a most appropriate model for a description of the effects leading to plasma nonuniformities caused by the excitation
of surface modes in low-pressure CCPs at VHF should be 1) electromagnetic, 2) self-consistent in terms of the coupled field-plasma description, 3) kinetic (at least for the electron component), and  4) nonlocal.    
The particle-in-cell (PIC) method commonly used for modeling the low-pressure CCPs combines the 2)-4) points. However, to the authors' knowledge, 
there have been no reports of a study of such effects with an electromagnetic PIC code (Note though that an energy-conserving code was successfully used in modeling of an ICP discharge \cite{mattei_2017}).  Although works utilized electrostatic PIC codes for investigations of
the surface mode excitation \cite{eremin_2016,eremin_2017a,eremin_2017b,wen_2017b}, they clearly lacked the electromagnetic aspect, which is important for a proper account of the symmetric mode excitation. This paper for the first time presents simulations of the CCP plasmas with large electrode radius, driven by VHF, and operated at low pressures, using an EM PIC/MCC code.

%Note that it is the uniformity of the flux and energy distribution for ions impinging on the electrodes that directly affects the plasma processing quality. Although the ion flux is commonly assumed to be closely related to the bulk plasma density, in the present work we demonstrate that they can exhibit quite different radial distributions.

%%%%%%%%%%%%%%%%%%%%%%
%\clearpage
\section{Numerical model} \label{sec2}

\subsection{Electromagnetic particle-in-cell (EM PIC) model}

To study the electromagnetic phenomena in CCP discharges, we have constructed a two-dimensional electromagnetic code ECCOPIC2M of the ECCOPIC code family developed in-house based on the fully implicit energy-conserving particle-in-cell method originally proposed by \cite{chen_2011,markidis_2011} and adapted for bounded plasmas \cite{Eremin2022}. Such an algorithm does not suffer from the finite grid instability when the cell size or time step exceeds the Debye length and plasma period, respectively \cite{Barnes2021}. The algorithm considers the time evolution of fields and particle orbits and consistently takes into account their interdependence at each time step, which enables energy conservation for a finite time step up to a controllable accuracy \cite{chen_2011,markidis_2011,chacon_2019,Eremin2022}. In addition, the chosen algorithm conserves the charge in the sense of the discretized charge continuity equation, obviating the need for solving Poisson's equation and improving the linear momentum conservation \cite{chen_2011}.

The energy- and charge-conserving particle-in-cell algorithm for bounded plasmas is verified for unmagnetized plasmas using the benchmark cases of \cite{turner_2013}, and for magnetized plasmas in the benchmark studies \cite{charoy_2019} and \cite{willafana_2021} (in the latter case output data of the implicit code was compared with those produced by the explicit code RUB, which was used in the reference). 
%In the present work we additionally employed a charge-conserving algorithm proposed in \cite{chen_2011}, which obviates the need to resort to any additional algorithms to enforce the Poisson equation and thus the charge conservation. 
Furthermore, we exploit a nonuniform mesh following the algorithm proposed in \cite{chacon_2013}. Above all, such a mesh makes it possible to effectively elongate the ``waveguide'' portion of the simulation domain in the axial direction in the lab coordinates, which is required for a more realistic description of the power coupling to the reactor. An additional coordinate transformation in the radial direction enables to strongly mitigate numerical difficulties at the radial axis. The logical mesh portion dedicated to the description of the plasma reactor itself is simultaneously kept large to ensure efficient usage of the computational grid.

\begin{figure}[h]
\centering
\includegraphics[width=12cm]{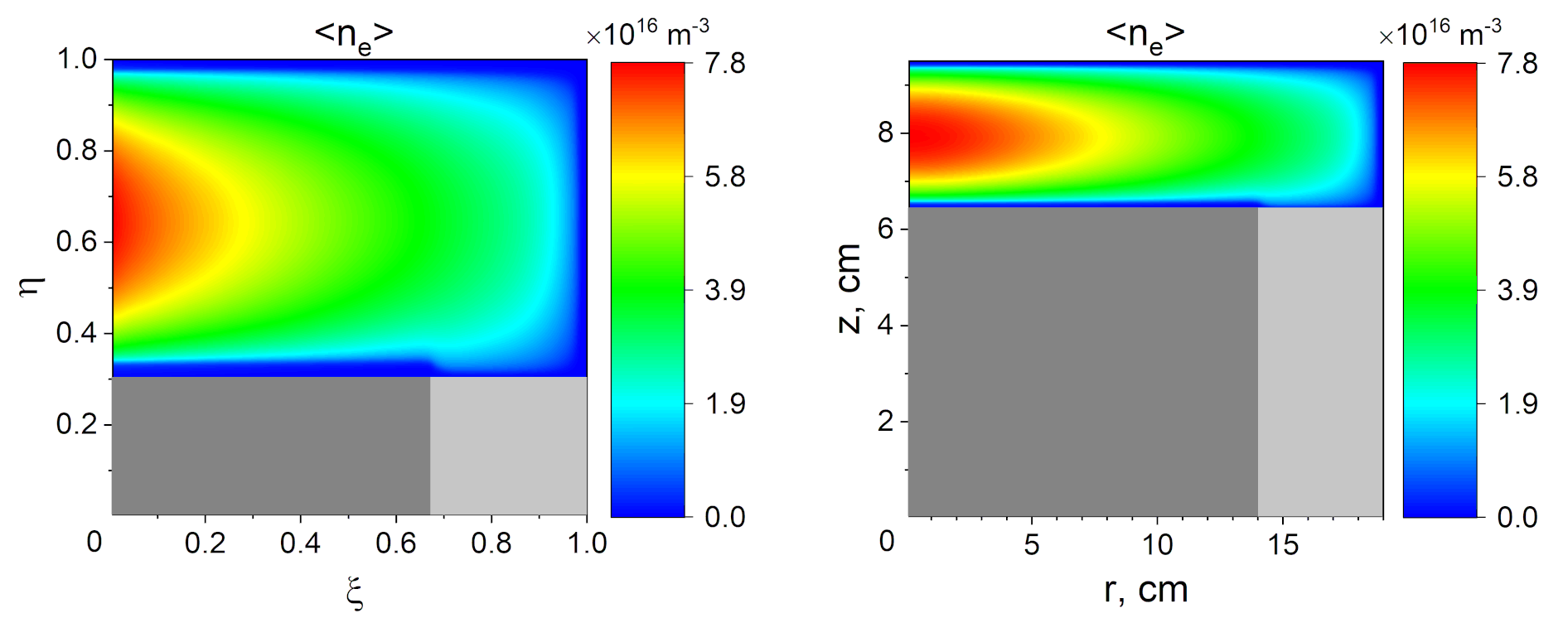}
\caption{The geometry of Testbench B in normalized logical coordinates (left) and in lab coordinates (right). The plasma shown in the reactor is modeled with EM PIC using $f=106$ MHz, $p=150$ mTorr, and $P_0=100$ W.}
\label{Fig1}
\end{figure}

In this work, we focus on modeling the experiment conducted on the Testbench B setup of \cite{sawada_2014} (see Fig.1, right). The discharge features a cylindrical CCP reactor driving plasmas at a frequency of $f=106$ MHz through the ``port'' from $r_p=14$ cm to $r_{max}=19$ cm, which is modeled as a dielectric with the permittivity of quartz. The reactor chamber is $3$ cm high, and the port length is $6.5$ cm. 
As mentioned above, such a relatively long port is chosen to reduce the effects caused by the possible electric field distortions generated at the boundary between the port and the main reactor chamber. Modeling the dielectric nature of the port material, any charged particles hitting it are not completely absorbed as is assumed to be the case for all metal surfaces, but rather stick to it. For the sake of simplicity, the secondary electron emission processes are neglected since the sheath voltages in simulations considered in Sec.\ref{sec4} are relatively low and no significant effects related to the secondary electrons are expected as it was found for similar conditions in \cite{Eremin2020}.
%leaving an investigation of their influence on the VHF CCPs for a %separate study. 
The discharge gas is taken to be argon at different pressures with a typical reaction set for low pressure (elastic, excitation, and ionization reactions for the electron-neutral collisions with the cross-sections are adopted from \cite{phelps_1999}, and the elastic and the charge-exchange reactions for the ion-neutral collisions with the cross-sections are taken from \cite{phelps_1994}). The corresponding collisions are implemented using the null-collision algorithm for GPUs \cite{mertmann_2011} with xorshift128 pseudorandom number generator \cite{Marsaglia_2003} randomly initialized for each thread.

Following the ansatz common to almost all of the previous works on the subject (e.g., \cite{lieberman_2002,sansonnens_2006,howling_2007}), we assume that the fields inside the CCP reactor are dominated by the TM mode, thus neglecting any azimuthal dependencies and retaining only the $E_r,E_z$, and $H_\theta$ field components. The latter is chosen to be the primary independent variable used for expressing all other quantities.
Since no external magnetic field was used in \cite{sawada_2014} (as is usually the case in regular CCP discharges), we neglect the influence of the magnetic field on the particle motion, since it is a second-order effect in terms of the field amplitude. 

The variable transformations are performed in each resolved coordinate separately, $r=r(\xi)$ and $z=z(\eta)$. In the logical coordinates $\xi,\eta$ the mesh is uniform. Proceeding in the spirit of
\cite{chacon_2013}, Faraday's and Ampere's laws in a curvilinear system of coordinates read
\begin{equation}
\left\{
\begin{array}{lll}
\frac{1}{J}\epsilon^{\alpha\beta\gamma}\frac{\partial}{\partial \beta} E_\gamma &=& - \mu \frac{\partial}{\partial t} H^\alpha \\
\frac{1}{J}\epsilon^{\alpha\beta\gamma}\frac{\partial}{\partial \beta} H_\gamma &=& j^\alpha + \left(\sigma+\epsilon \frac{\partial}{\partial t}\right) E^\alpha,
\end{array}
\right. \label{eq2_1}
\end{equation}
where $J$ is the transformation's Jacobian $J=r J_r J_z$ with $J_r = dr/d\xi$ and $J_z=dz/d\eta$, $\epsilon$ with the superscripts is the Levi-Civita symbol, the quantities with the superscripts (subscripts) denoting contravariant (covariant) field components related by the metric tensor \cite{chacon_2013}, and $\mu$, $\epsilon$, and $\sigma$ are the material magnetic permeability, dielectric permittivity, and electric conductivity, respectively. Based on our experience, utilization of the large values for the latter is preferential for modeling of conducting materials such as metals in electromagnetic simulations, whereas in electrostatic simulations one can as well set the dielectric permittivity to large values \cite{Eremin2022}. 

Picking the field components relevant to the problem in question, one obtains
\begin{equation}
\left\{
\begin{array}{lll}
\mu \frac{\partial}{\partial t} H^\theta &=& \frac{1}{J}\left(\frac{\partial}{\partial\xi} E_\eta - \frac{\partial}{\partial \eta} E_\xi\right) \\
\left(\sigma+\epsilon \frac{\partial}{\partial t}\right) E^\xi &=& -j^\xi - \frac{1}{J}\frac{\partial}{\partial\eta} H_\theta \\
\left(\sigma+\epsilon \frac{\partial}{\partial t}\right) E^\eta &=& -j^\eta + \frac{1}{J}\frac{\partial}{\partial \xi} H_\theta,
\end{array}
\right. \label{eq2_2}
\end{equation}
and the field components are discretized according to the Yee scheme \cite{yee_1966}.
To complete the field equations, one needs to specify a discretized form for the current density, which in the considered case of a two-dimensional problem is provided by \cite{chacon_2013}
\begin{equation}
\begin{array}{lll}
j^{\xi} &=& \frac{1}{J\Delta\xi\Delta\eta}\sum\limits_p q_p  v_{rp} J_r(\xi_p) S_{m-1}(\xi_p-\xi_{i+1/2}) S_m(\eta_p-\eta_j) \\ 
j^{\eta} &=& \frac{1}{J\Delta\xi\Delta\eta}\sum\limits_p q_p v_{zp} J_z(\eta_p) S_{m}(\xi_p-\xi_i) S_{m-1}(\eta_p-\eta_{j+1/2}),
\end{array} \label{eq2_3}
\end{equation}
where $\Delta \xi$ and $\Delta \eta$ denote the cell sizes of the logical grid, $S_m$ is the b-spline of order $m$ \cite{chen_2011}, $p$ subscript denotes contribution from the $p$th particle, and $i$ or $j$ denote the vertices of the corresponding computational mesh. 

A new technique is used to mitigate problems related to the numerical integration of the field equations at the radial axis, which takes advantage of the nonuniformity of the radial grid. Namely, the transformation of the radial coordinate is chosen in such a way that $r \propto \sqrt{\xi}$ for just a few points close to the axis, while $r \propto \xi$ for the rest of the grid. In this case, $rJ_r$ goes to a constant value at the axis, and no numerical complications arise \cite{eremin_2021b}. This method enables an optimal use of the available logical gridpoints in the radial direction for the reactor resolution since big radial cells are only used close to the axis, and the rest of the computational domain is well resolved. Further, if the particle weight is simultaneously taken to be proportional to $rJ_r$, it also remains finite for the particles initially located close to the axis. Therefore, the well-known problem of having superparticles with varying weight in cylindrical geometry, where in the course of simulations the heavyweight particles from the radial edge tend to replace the extremely lightweight particles at the axis and vice versa, is also strongly mitigated. Because of a moderate ratio of the largest to the smallest superparticle weight, we did not use the adaptive particle management techniques.
%It also removes the need for adaptive particle management techniques, which could disturb particle velocity distributions potentially leading to unphysical effects.

In order to make the model self-consistent, one needs to specify the time evolution of particle orbits. It is governed by the following equations in cylindrical coordinates \cite{delzanno_2013b},
\begin{equation}
\left\{
\begin{array}{rll}
m_s\frac{dv_r}{dt} &=& q_sE_r + m_s\frac{v_\theta^2}{r} \\
m_s\frac{dv_\theta}{dt} &=& - m_s\frac{v_r v_\theta}{r} \\
m_s\frac{dv_z}{dt} &=& q_s E_z \\
\frac{d\xi}{dt} &=& \frac{v_r}{J_r}\\
\frac{d\eta}{dt} &=& \frac{v_z}{J_z} , 
\end{array}
\right.   \label{eq2_4}
\end{equation}
where particle velocities are $v_r = dr/dt$ and $v_z=dz/dt$ and the positions
are evolved in the logical coordinates, which makes it easy to track particles \cite{chacon_2013,wang_1999,delzanno_2013}. Eqs. (\ref{eq2_1}-\ref{eq2_4}) must be discretized in time, whereby time derivatives should be replaced by $(\partial_t f)^{n+1/2} = (f^{n+1}-f^n)/\Delta t$, and all other quantities should be taken at the mid-level between the new and the old time levels, $f^{n+1/2} = (f^n+f^{n+1})/2$. To integrate particle orbits in time, Eq. (\ref{eq2_4}) is used directly (in contrast to the typical use of the Boris's algorithm in cylindrical coordinates, e.g., \cite{mattei_2017}). This simplifies particle tracking and detection of particle crossings of cell boundaries, which is needed to ensure the charge conservation \cite{chen_2011}. The use of such an algorithm does not lead to noticeable errors or deviations from the results obtained with the standard Boris algorithm as was demonstrated, for example, in \cite{charoy_2019}.

To complete the numerical EM model, one also has to specify how energy is supplied to the system. This is done by prescribing an incoming TEM wave at the dielectric port located at the bottom of the computational domain. However, one has to be careful 
and avoid using the ``hard source'' technique fixing the azimuthal magnetic field component to be equal to the incoming wave only. This does not allow the outgoing wave produced by various reflections to leave the computational domain, creating spurious solutions or leading to diverging simulations. A method to overcome this difficulty is to formulate the boundary condition in the form of the Mur condition for the outgoing TEM wave $(J_z^{-1}\partial/\partial \eta + (\sqrt{\epsilon\mu}/c) \partial/\partial t)H_\theta^- = 0$ \cite{mur_1981}. Then, recalling that $H_\theta^-=H_\theta - H_\theta^+$ with $H_\theta$ the total magnetic field intensity and $H_\theta^+$ the incoming wave, one
can write the boundary condition for $H_\theta$ at the port as \cite{alvarez_2007,jimenez-diaz_2011,rahimi_2014}
\begin{equation}
\left(\frac{1}{J_z}\frac{\partial}{\partial \eta}  - \frac{\sqrt{\epsilon\mu}}{c}\frac{\partial}{\partial t} \right) H_\theta = S \label{eq2_5}
\end{equation}
with the source $S$ specified by the prescribed incoming wave, $S = 2 (\sqrt{\epsilon\mu}/c)\partial H^+_\theta/\partial t$. In this way the outgoing wave is free to leave the computational domain virtually without reflections at the port. This also solves the potential problem of the radiative noise accumulation, which is undamped to the the algorithm's energy conserving properties \cite{markidis_2011} and can compromise a simulation if electromagnetic waves are not allowed to exit the computational domain. Since it is not the case in the present method, it allows to consistently capture effects related to electromagnetic waves at very high frequencies and large phase velocities without accumulating the radiative noise. Should such effects not be important, another possible option is to use an energy-conserving algorithm based on the Darwin approximation \cite{chen_2014,chen_2015,chacon_2016}.    

To appropriately account for a possible sheath asymmetry, a description of an external network having a blocking capacitor sustaining a dc self-bias is needed. It requires knowing which part of the total current on the reactor's side is generated by the incoming wave. In case when one has a matching network, which guarantees that there is no reflected wave, there is only the incoming wave, and all current is generated by it. In this case one can proceed similar to the electrostatic case \cite{Eremin2022}. However, we decided to reserve the effects associated with complex external networks \cite{yamazawa_2015,schmidt_2018,schmidt_2019} for a separate study and consider here only a simple external network with a single blocking capacitor of capacitance $C$. In this case of an unmatched reactor load, the total plasma impedance is dominated by the capacitive sheaths. It causes only a relatively small fraction of the incoming power to be absorbed by the plasma bulk, and a large fraction of the power produced by the generator to be reflected back. Since the port's axial length is much smaller than the wavelength at even $106$ MHz in quartz (around $1.5$ m), the reflected wave has practically the same phase. Therefore, we use the same procedure as in the electrostatic case here as well. Namely, we first calculate the powered electrode potential $\phi^{n+1}$ needed to match currents from the reactor and the external network sides at the powered electrode's surface \cite{verboncoeur_1993,vahedi_1997,Eremin2022},
\begin{equation}
2\pi \int\limits_0^{\xi_p} \left(\frac{\epsilon_0 rJ_r }{J_z}\left(\frac{\partial E_\eta}{\partial t}\right)^{n+1/2}+j^{\eta,n+1/2}\right)d\xi = \frac{C(V^{n+1}-\phi^{n+1})-q^n}{\Delta t} \label{eq2_6}
\end{equation}
with $V$ the generator voltage source, $j^\eta$ the axial contravariant component of the plasma current (see Eq.(\ref{eq2_3})) at the electrode, and $q^n$ the capacitor charge at the previous time level. Using the reasoning given above and assuming that the potential is generated by the incoming and the outgoing waves of approximately equal amplitudes and phases, one can then assume that the part of the potential generated by the incoming wave only should be $\phi^{+,n+1} \approx \phi^{n+1}/2$. Finally, one can calculate the source $S$ in Eq.(\ref{eq2_5}) by using the appropriate relation between $\phi$ and $H_\theta$ in a TEM mode, which yields
\begin{equation}
S^{n+1/2} = - \frac{\epsilon}{c}\left(r_p\ln\frac{r_{max}}{r_p}\right)^{-1}\frac{\phi^{n+1}-\phi^n}{\Delta t} \label{eq2_7}
\end{equation}
with $\phi^{n+1}$ calculated from Eq.(\ref{eq2_6}) at this time level, and $\phi^n$ being the value calculated before at the previous time level. The voltage source amplitude $V$ is chosen to match a prescribed power value $P_0$ using the approach described in \cite{eremin_2016b}.
We found that the suggested procedure yields the same voltage form at the powered electrode (in terms of the amplitude, phase, and the dc self-bias) as in the purely electrostatic analog described in the next section in the appropriate limit of a large speed of light (see Section \ref{subsec3c}). Finally, it is worth noting that this algorithm allows to implement a general external network by modifying the right-hand side of Eq.(\ref{eq2_6}) as suggested in \cite{Eremin2022}. An alternative approach (though functional solely for an external circuit with a blocking capacitor) could be to gradually adjust the dc bias until the time-averaged electron and ion fluxes to the powered electrode match.
%(e.g., [REFERENCE]).   

The overall algorithm built this way conserves energy to a desired prescribed accuracy in the sense of the Poynting theorem \cite{brackbill_2015,chacon_2019}.
Since it results in a system of coupled equations that must be solved simultaneously, it represents an implicit algorithm. The equations are typically solved using an iterative algorithm, such as the Jacobian-free Krylov-Newton method \cite{chen_2011,markidis_2011,mattei_2017} or the Picard method \cite{taitano_2013,Eremin2022} and do not lead to numerical problems even if the Debye length or the time interval it takes a light wave to cross a cell are not resolved \cite{chen_2011,markidis_2011}. To reduce computational time, EM PIC and ES PIC (see next section) are parallelized on GPU using a two-dimensional generalization of an algorithm similar to the one described in \cite{mertmann_2011}.

\subsection{Electrostatic particle-in-cell (ES PIC) model}

To differentiate the electromagnetic effects from the effects
that can be observed already in the electrostatic approximation (note, for example, that the antisymmetric mode is basically the same in the electrostatic approximation in the case when the skin depth is much larger than the electrode gap), the electrostatic version ECCOPIC2S is used \cite{eremin_2021b}. It is different from the electromagnetic version described in the previous subsection in that the electric field contains only the potential part, described by Poisson's equation
\begin{equation}
\frac{1}{rJ_r}\frac{\partial}{\partial \xi}\frac{\epsilon r}{J_r}\frac{\partial}{\partial \xi} \phi  + \frac{1}{J_z}\frac{\partial}{\partial \eta}\frac{\epsilon}{J_z}\frac{\partial}{\partial \eta}\phi = -e(n_i-n_e)
\label{eq2_8}
\end{equation}
along with $E_\xi = -\partial \phi/\partial \xi$ and $E_\eta = -\partial \phi/\partial \eta$. To specify the boundary condition for Eq.(\ref{eq2_5}) a usual assumption that $\phi$ is set to an appropriate potential at each surface except the dielectric spacer is exploited. For the latter $E_r=0$ at the dielectric spacer at the bottom of the waveguide is employed, which yields a fixed radial profile for the electrostatic potential $\{\phi(r) = \phi_0 \ln(r/r_{max})/\ln(r_p/r_{max})\,,\, r_p\leq r \leq r_{max}\}$ with
$\phi_0$ the powered electrode potential. This represents a Dirichlet boundary condition at the port.  

When the external network is used, the potential is calculated from the need to satisfy Kirchhoff's laws 
at the driven electrode \cite{verboncoeur_1993,vahedi_1997,Eremin2022}. If the external network has just a single capacitor, one arrives at the same equation as Eq.(\ref{eq2_6}), which in the case of an implicit code and the accompanying iterative solution procedure can be regarded as an equation for the next iteration of $\phi^{n+1}$ on the right-hand side, based on $\partial E_\eta^{n+1/2}/\partial t$ calculated from the previous iteration of $E_\eta^{n+1}$ and its value on the previous time level $E_\eta^n$ along with the previous iteration of the plasma current $j^{\eta,n+1/2}$ \cite{Eremin2022}.

The corresponding ES PIC utilizing the Poisson equation yielding the electrostatic potential $\phi$ conserves energy as long as the charge conservation is ensured, which is the case in the electrostatic ECCOPIC2S. It is worth mentioning that charge conservation is not necessary for energy conservation, although in this case, one must update the potential using the divergence of Ampere's law rather than the Poisson equation \cite{chen_2011}.

%%%%%%%%%%%%%%%%%%%%%%
%\clearpage
\section{Model verification} \label{sec3}

After building a new model it is important to verify that it yields numerical surface mode dispersion relations 
complying with the corresponding analytical theories (see Sec.\ref{sec1}) and is consistent with the well-tested electrostatic version of the code in the appropriate limits (a small driving frequency or $c\to \infty$). To this goal, we conduct three different verification tests described in this section.

\subsection{Surface mode dispersion relations} \label{subsec3a}

This test is intended to prove that the electromagnetic effects are correctly implemented by checking if the dispersion relation extracted from the particle-in-cell code agrees with the expression derived for a simple case of uniform plasma with symmetric sheaths \cite{lieberman_2002,sansonnens_2006}. To this end, ions are artificially fixed at initial positions yielding a uniform density profile with $n_0 = 10^{16} \, {\rm m}^{-3}$. The simulation is run for some time with a sinusoidal voltage having an amplitude of $100$ V. The geometry is kept the same as for the main simulations. In this test, the external network with a blocking capacitor is not used, and collisions are not included. This results in the formation of quasi-symmetrical electron-depleted sheaths on each electrode (except for the sheath at the dielectric spacer) of $1.8$ mm width. After the sheaths are well formed, the main driving voltage is turned off, and the plasma discharge is excited instead with a sawtooth-like voltage having the form 
$$
V_{ex}(t) = A \sum\limits_{n=1}^N \frac{\sin(2\pi n f t)}{n}
$$
with an amplitude $A$ small enough not to create a large perturbation of the sheaths, but large enough to produce a noticeable broad-spectrum response from the plasma-filled reactor. In this case, $A=1$ V, $N=20$, and $f=106$ MHz are chosen. The response of the axial electric field $E_z$ is then analyzed at a fixed axial distance from the electrode of $1$ mm 
using the Fourier-Bessel decomposition along the radial coordinate (see \ref{Appendix_Fourier_Bessel}) and Fourier decomposition in time. The spectrum for the axial electric field obtained in such a way should manifest peaks at values given by $\omega(k_m)$ with $\omega(k)$ the dispersion curve and $k_m$ is the radial wavenumber corresponding to the $m$th mode, which in case of rectangular geometry for the symmetric modes is given by $\pi(2m-1)/L$ \cite{eremin_2017b}, and in case of cylindrical geometry equals to $k_m = \chi_{0m}/R$ with $\chi_{0m}$ the $m$th zero of the Bessel function $J_0(x)$ \cite{lieberman_2015}. Similarly, for the antisymmetric modes it should yield $k_m = \chi_{1m}/R$ with $\chi_{1m}$ the $m$th zero of the Bessel function $J_1(x)$. 

\begin{figure}[h]
\centering
\includegraphics[width=12cm]{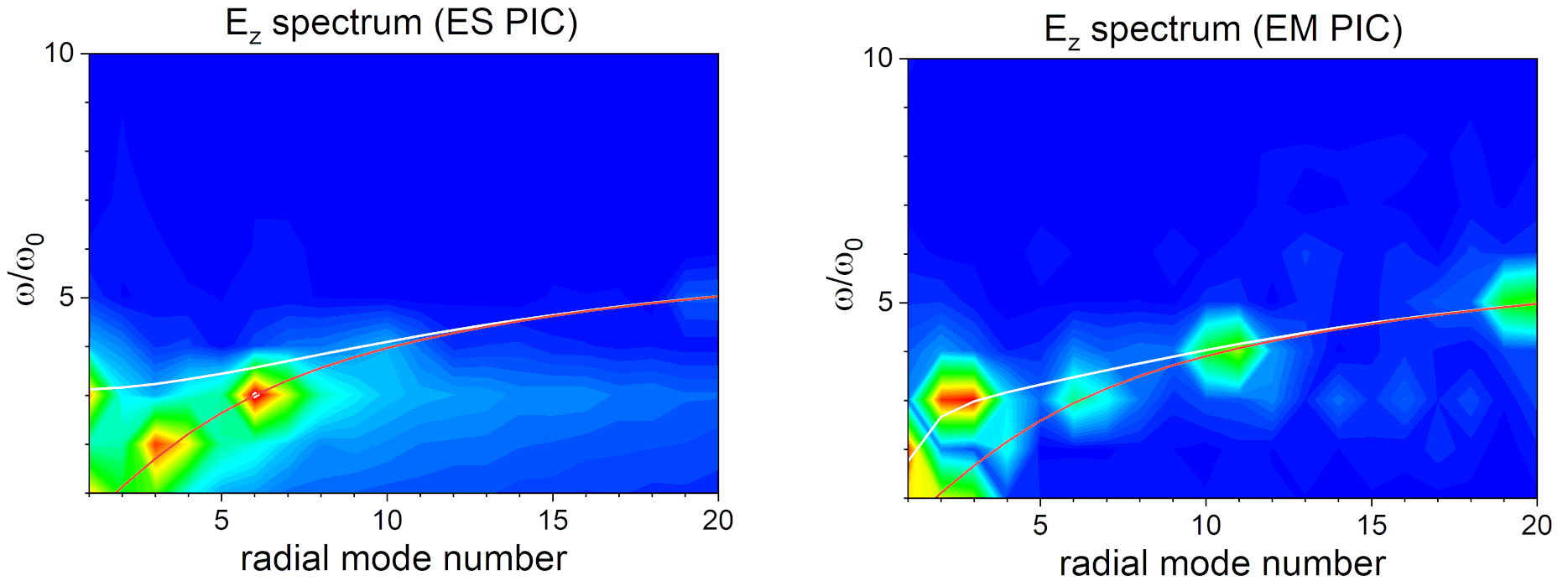}
\caption{Fourier-Bessel spectrum of the axial electric sampled at an axial location in one of the sheaths for a model problem (see description in the text).}
\label{Fig2}
\end{figure}

To track the change between the electrostatic and the electromagnetic descriptions, the above procedure is performed for both. The resulting dispersion relations can be seen in Fig.~\ref{Fig2}. The analysis of the axial electric field picks up strong signals for both the symmetric and antisymmetric mode types, since the data are sampled in one of the sheaths.
For comparison, analytical curves for the corresponding symmetric and antisymmetric curves are superimposed on the spectra extracted from PIC simulations. 

The corresponding curves are provided by solutions of the following equations \cite{bowers_2001,lieberman_2002,sansonnens_2006}:
\begin{equation}
\left\{
\begin{array}{lll}
\beta\coth(\alpha s) &=& -\alpha\epsilon_p \coth(\beta d) \,,\, {\rm symmetric} \\ 
\beta\coth(\alpha s) &=& -\alpha \epsilon_p \tanh(\beta d) \,,\, {\rm antisymmetric}
\end{array} \right.
\label{eq3_1}    
\end{equation}
in the electromagnetic description (here $\alpha = (k^2-\omega_{pe}^2/c^2)^{1/2}$, $\beta=(k^2-\omega_{pe}^2\epsilon_p/c^2)^{1/2}$, and the relative plasma dielectric permittivity $\epsilon_p = 1 - \omega_{pe}^2/\omega^2$) and
\begin{equation}
\left\{
\begin{array}{lll}
\omega &=& \omega_{pe}\left( \frac{\coth (kd)}{\coth (ks) + \coth(kd)}\right)^{1/2} \,,\, {\rm symmetric} \\ 
\omega &=& \omega_{pe}\left( \frac{\tanh (kd)}{\coth (ks) + \tanh(kd)}\right)^{1/2} \,,\, {\rm antisymmetric}
\end{array} \right.
\label{eq3_2}    
\end{equation}
in the electrostatic approximation obtained through $c\to\infty$. Note that in the latter case the dispersion relation for the symmetric mode has a cutoff frequency at the plasma series resonance frequency, $\omega_{PSR} = \omega_{pe}\sqrt{s/l}$, whereas in the electromagnetic case it has the expected knee in the long-wavelength part \cite{bowers_2001,eremin_2016,eremin_2017a,eremin_2017b,wen_2017b}. 

Fig.\ref{Fig2} clearly shows that such a transition between the expected behavior in the electromagnetic and the electrostatic cases for the symmetric mode's dispersion relation is well reproduced by the electromagnetic PIC code. Whereas the antisymmetric mode's dispersion relation is hardly changed between the EM and ES versions, which is to be expected when the skin depth is much larger than the electrode gap \cite{eremin_2017a} (for the considered parameters $\delta=c/\omega_{pe}\approx 5.3$ cm $\gg$ $d = 1.5$ cm), the dispersion curve for the symmetric mode in the electromagnetic case obtained from PIC data follows the analytical dispersion curve going down at $m=1$, where $m$ is the radial mode number. Good agreement between the analytical prediction and the PIC results is also seen for the symmetric mode in the electrostatic approximation, where the cutoff of the spectrum at the predicted plasma series resonance frequency can be observed. 
%Whereas the ES PIC spectrum does show excitation of higher radial mode numbers, %the dominance of the low radial mode numbers is much stronger compared to the EM %PIC.  

\subsection{Comparison of ES PIC and EM PIC results for f = 13.56 MHz} \label{subsec3b}

\begin{figure}[h]
\centering
\includegraphics[width=12cm]{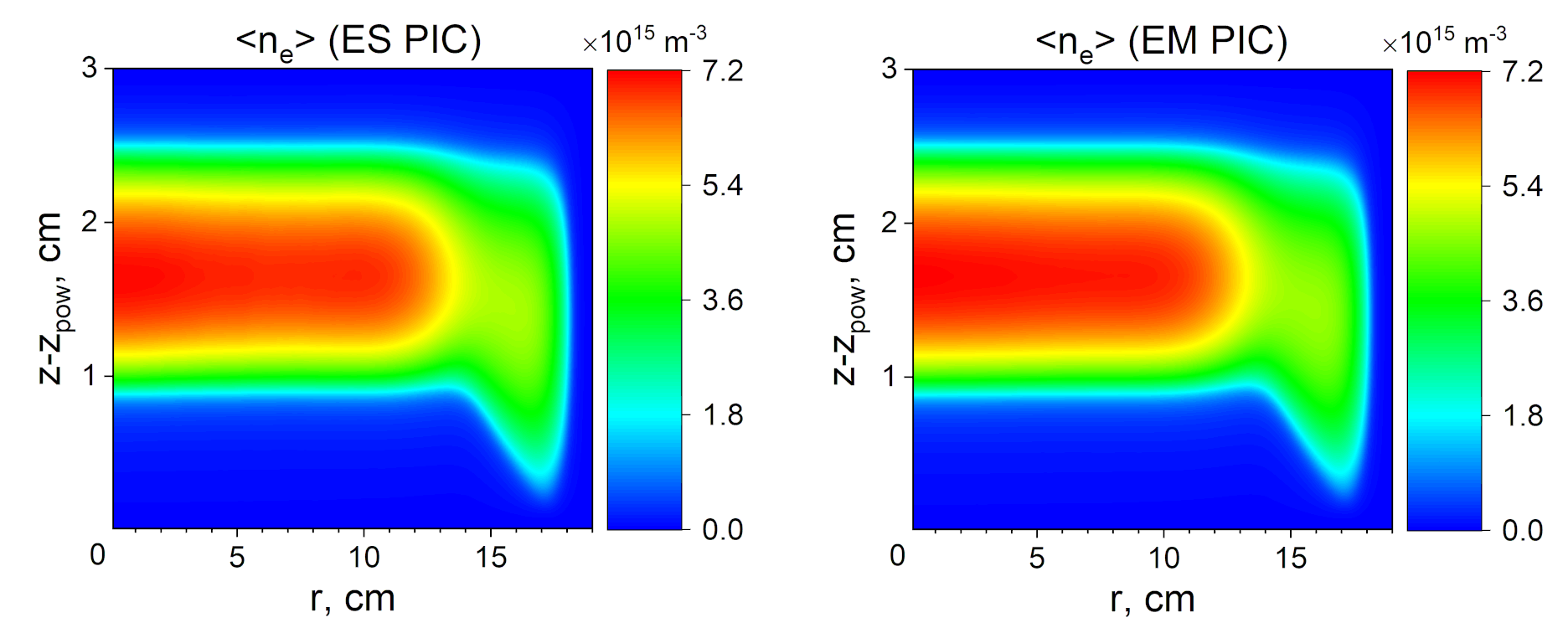}
\caption{Comparison of the plasma density profiles calculated with the electrostatic and electromagnetic versions of the PIC/MCC code for $f=13.56$ MHz. The axial coordinate is counted from the powered electrode's axial location, $z=z_{pow}$.}
\label{Fig3}
\end{figure}

In the second verification test, we compare the EM and ES PIC simulation results for a low-frequency CCP discharge where no significant electromagnetic effects are anticipated. To this end, we use $f=13.56$ MHz and $P_0=25$ W. The resulting period-averaged electron density profile is indeed very similar in both simulations, see Fig.\ref{Fig3}. In this case, the discharge develops a large negative self-bias to match the period-averaged electron and ion fluxes to the powered electrode. It can be observed that the new approach aimed at including the external network with a blocking capacitor in EM PIC proposed in the previous section correctly reproduces the self-bias seen in the results provided by the ES PIC.

\subsection{Transition of the electromagnetic results into electrostatic ones with the increase of c} \label{subsec3c}

\begin{figure}[h]
\centering
\includegraphics[width=15.5cm]{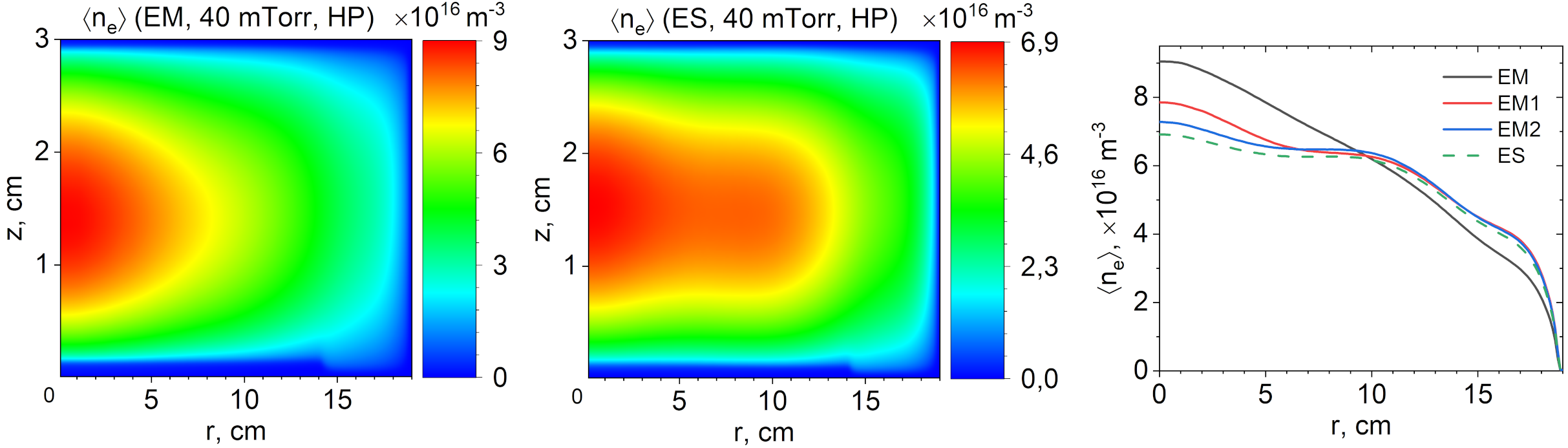}
\caption{The transition of the mid-plane electron density profile obtained with the fully EM PIC code to the results provided by the ES PIC version as the speed of light in the EM PIC is artificially increased. Calculations are made for the HP case at $40$ mTorr (see \ref{sec4.1} for the description of this case).}
\label{Fig_elDen_HP_40mTorr_EM_to_ES}
\end{figure}

The final verification test checks if the transition from the electromagnetic to the electrostatic description as the speed of light in it is artificially increased \cite{bowers_2001} (which is effectively made by increasing the magnetic permeability of vacuum $\mu_0$) is reproduced by the EM PIC. Fig.\ref{Fig_elDen_HP_40mTorr_EM_to_ES} demonstrates such a transition for the period-averaged electron density obtained during the converged state from the electromagnetic case (left plot) to the purely electrostatic one (middle plot), the latter having been obtained with ES PIC. The transition itself can be observed on the radial profile of the density sampled at the reactor midplane (right plot), where EMn stands for runs with $c_*=10^n c$, where $n=1,2$ and $c_*$ is the speed of light used in EM PIC. One can see that even when $c_*=10 c$, the electron density profile comes close to the electrostatic solution for most of the discharge except in the vicinity of the radial center. However, the fine details of the profile continue changing as the speed of light is further increased and even for $c_*=10^2 c$ there is a slight (although small) discrepancy between the EM and ES PIC.

%%%%%%%%%%%%%%%%%%%%%%
%\clearpage
\section{Results} \label{sec4}

\begin{figure}[h]
\centering
\includegraphics[width=12cm]{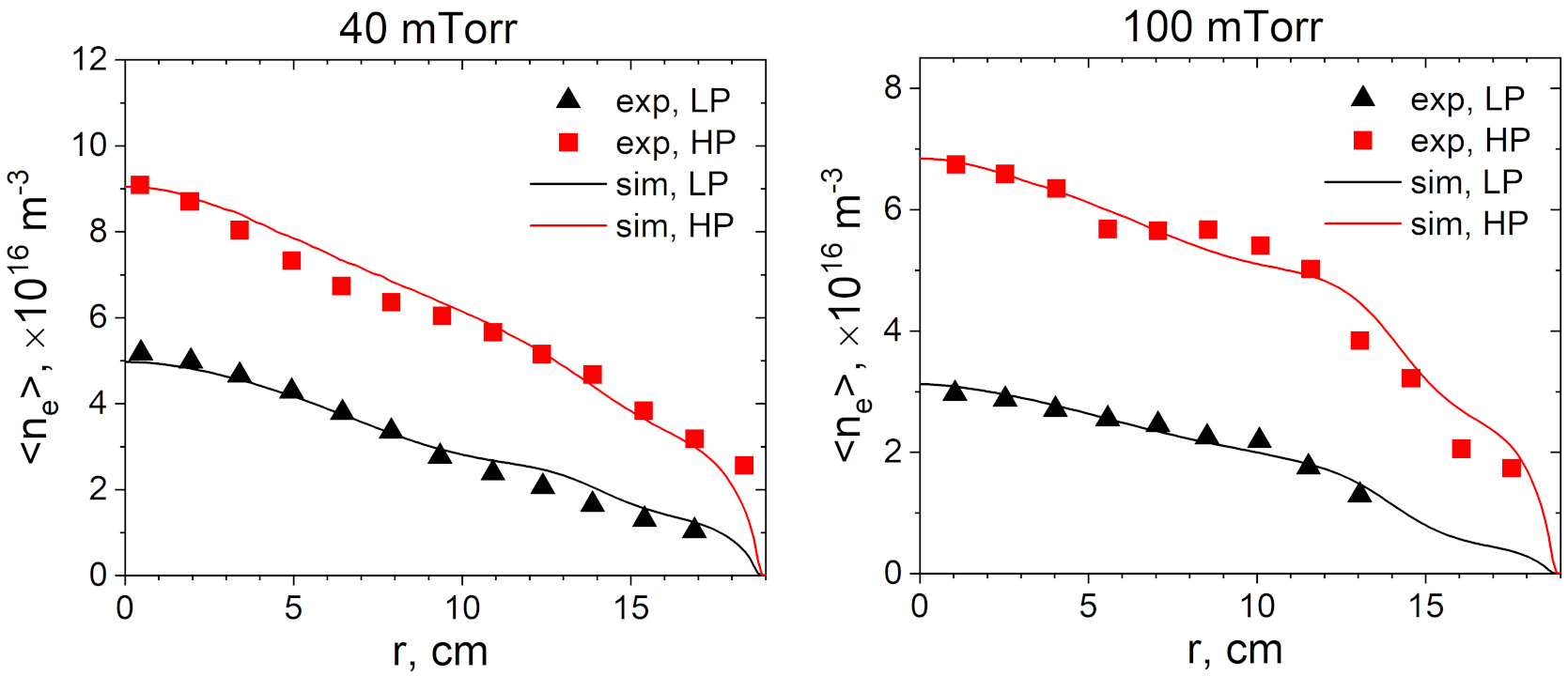}
\caption{Comparison of the numerical results obtained with the ECCOPIC2M code based on the energy- and charge-conserving EM PIC/MCC scheme and of the experimental data measured on the Testbench B CCP reactor and reported in \cite{sawada_2014}.}
\label{Fig_ne_in_validation_cases}
\end{figure}

\subsection{Validation of the EM PIC code with the time-averaged plasma density profiles} \label{sec4.1}

Fig.~\ref{Fig_ne_in_validation_cases} shows the results of the EM PIC simulations performed for the cases studied experimentally on the Testbench B reactor in \cite{sawada_2014}. To perform the comparison, the absorbed power in the simulations is adjusted to match the central plasma density values between the simulations and the experimental data. Since the influence of secondary electrons on discharge uniformity is found to be moderate for the parameter range considered \cite{Eremin2020} (note, however, that secondary electrons might be essential under different conditions \cite{Lieberman_2022}), the secondary electron emission is not included in the present work for the sake of simplicity and is reserved for a later study. The absorbed power values obtained from the matching procedure are equal to $37$ and $84$ W for the $40$ mTorr simulations and $15$ and $50$ W for the $100$ mTorr simulations, respectively. In the following, we will refer to the lower-power cases at each pressure as LP, and to the higher-power cases as HP. Although these values are arguably different from those quoted in the experimental work \cite{sawada_2014} ($100$ and $400$ W), they
are in the same range as those obtained in a recent simulation study utilizing a different model \cite{liu_2022}. This can be explained by large power losses in the experiment, which are hard to account for. Hence, matching the plasma density central values between the experiment and simulations for a validation study is more reasonable than matching the power values. 

The radial plasma profiles are governed by the discharge physics, which involves a complex interplay between the neutral gas ionization and the plasma transport. Therefore, if the radial profiles agree between simulations and experimental data, one can conclude that the basic physics is captured well by the model. Fig.~\ref{Fig_ne_in_validation_cases} demonstrates that it is indeed the case. By looking at Fig.~\ref{Fig_elDen_HP_40mTorr_EM_to_ES} obtained for parameters similar to the HP case at $40$ mTorr, one can see that the electromagnetic description is essential for achieving a good agreement between the experimental and simulation results. 
%As we argue later, it leads to a more stronger excitation of the symmetric mode, which affects the radial ionization profile stronger compared to the asymmetric mode.

\begin{figure}[h]
\centering
\includegraphics[width=12cm]{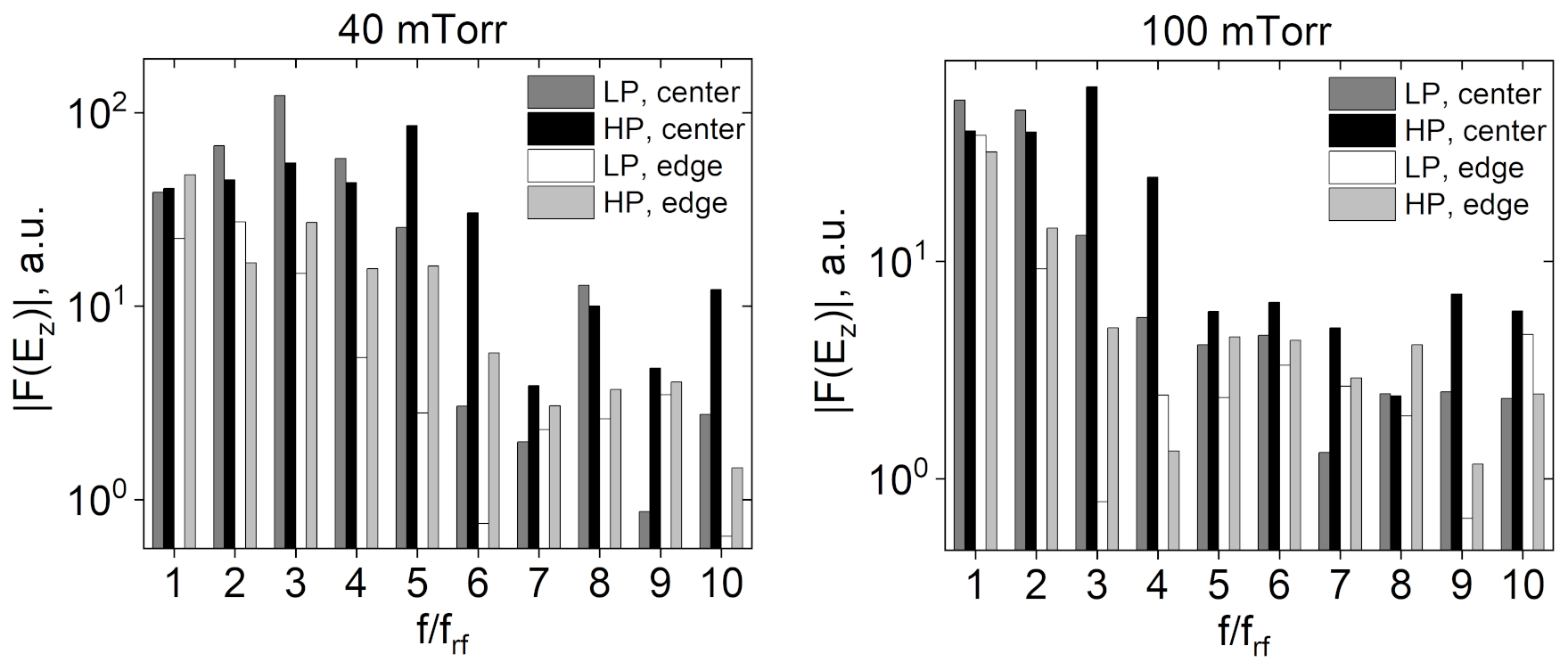}
\caption{Fourier amplitudes of the axial electric field harmonics sampled at the reactor mid-plane for the cases shown in Fig.~\ref{Fig_ne_in_validation_cases}.}
\label{Fig6}
\end{figure}

\begin{figure}[h]
\centering
\includegraphics[width=12cm]{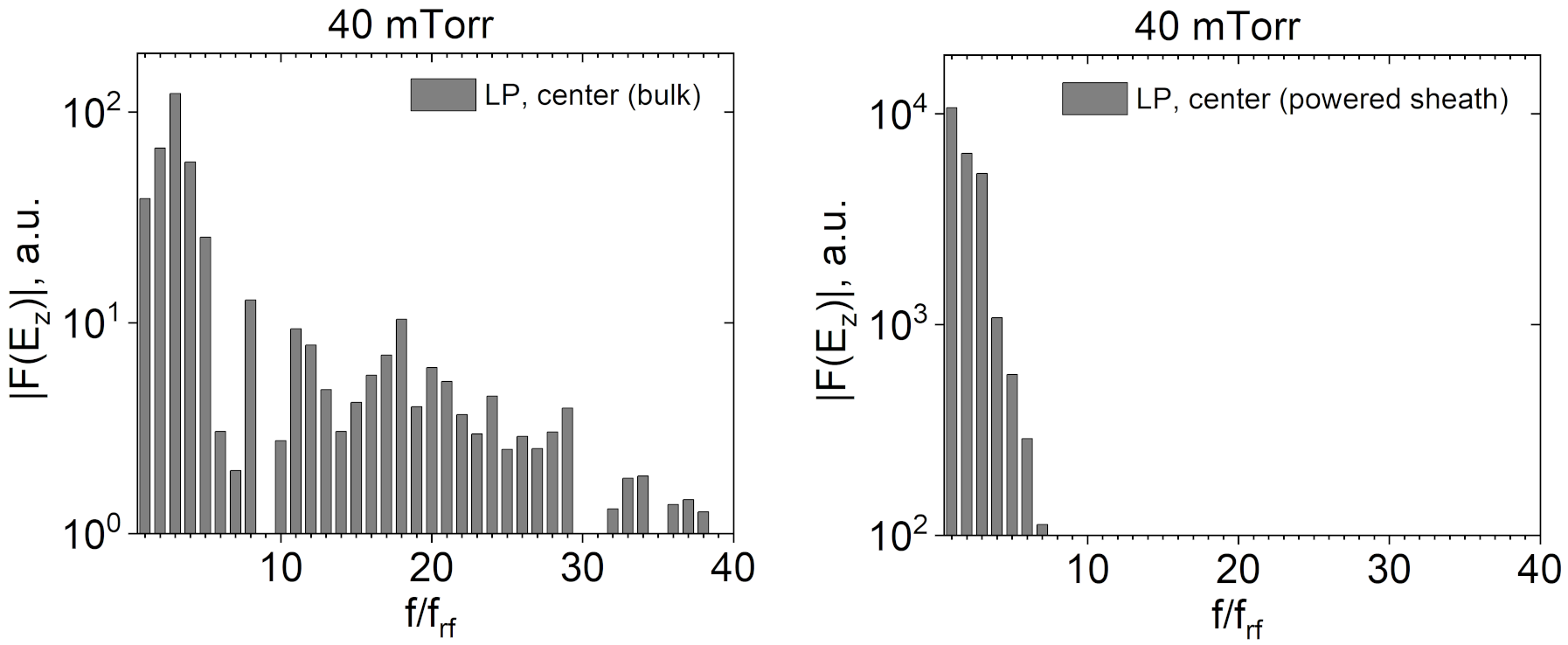}
\caption{Fourier amplitudes of the axial electric field harmonics sampled at the reactor mid-plane and close to the power electrode for the LP $40$ mTorr case.}
\label{Fig7}
\end{figure}

Another result obtained experimentally in \cite{sawada_2014} is the detection of a broad spectrum of the axial electric field harmonics at the reactor mid-plane (see Fig.~5 in the reference). A direct comparison with the simulation data is complicated because the measurements in \cite{sawada_2014} were not calibrated and were provided only to indicate main trends. In this context, our simulations also exhibit many harmonics of the fundamental excitation frequency in the Fourier spectrum of the axial field sampled at the mid-plane close to the center $r=1$ cm and close to the edge $r=17$ cm, see Fig.~\ref{Fig6}. Similar to \cite{sawada_2014}, there are more strongly excited higher harmonics ($1-6$ for $40$ mTorr and $1-4$ for $100$ mTorr) close to the center than close to the edge, where the spectrum rapidly falls with the harmonic number. The dominant harmonic range at the center becomes smaller for higher pressure (especially for the LP case), which also agrees with observations done in other simulation studies. 

\subsection{Further numerical investigation of the validation cases}

It is also interesting to take a step beyond the experimental limitations and to
look at a wider range of the excited harmonics compared to the spectrum shown in \cite{sawada_2014}, which was truncated at the $9$-th harmonic. Fig.~\ref{Fig7}a demonstrates the spectrum of Fourier amplitudes up to the $40$-th harmonic of the axial electric field sampled at the mid-plane close to the radial center for the LP $40$ mTorr case. One can see another peak at the $18$-th harmonic in the spectrum. It    
can be attributed to the excitation of the parallel resonance \cite{Ku_1998a,Ku_1998b} in the bulk plasma by energetic electron beams generated via the sheath motion \cite{wilczek_2016}. Using $n_e=5\times 10^{16}$ m$^{-3}$, one can estimate the ratio between the plasma frequency and the angular excitation frequency to be $\omega_{pe}/\omega\approx 19$, which is close to the observed peak. A broad spectrum of the excited harmonics in CCP reactors is not uncommon (see, e.g., Fig.~3 in \cite{wilczek_2018}). However, the corresponding high harmonics have a relatively small contribution to the absorbed power for the considered discharge parameters, which can be seen from the corresponding power spectrum (not shown). Furthermore, the high-frequency part is not observed in the spectrum of the axial electric field harmonics excited in the sheaths (Fig.~\ref{Fig7}b shows such a spectrum for the powered sheath), which is also expected for parallel resonance. 

\begin{figure}[h]
\centering
\includegraphics[width=12cm]{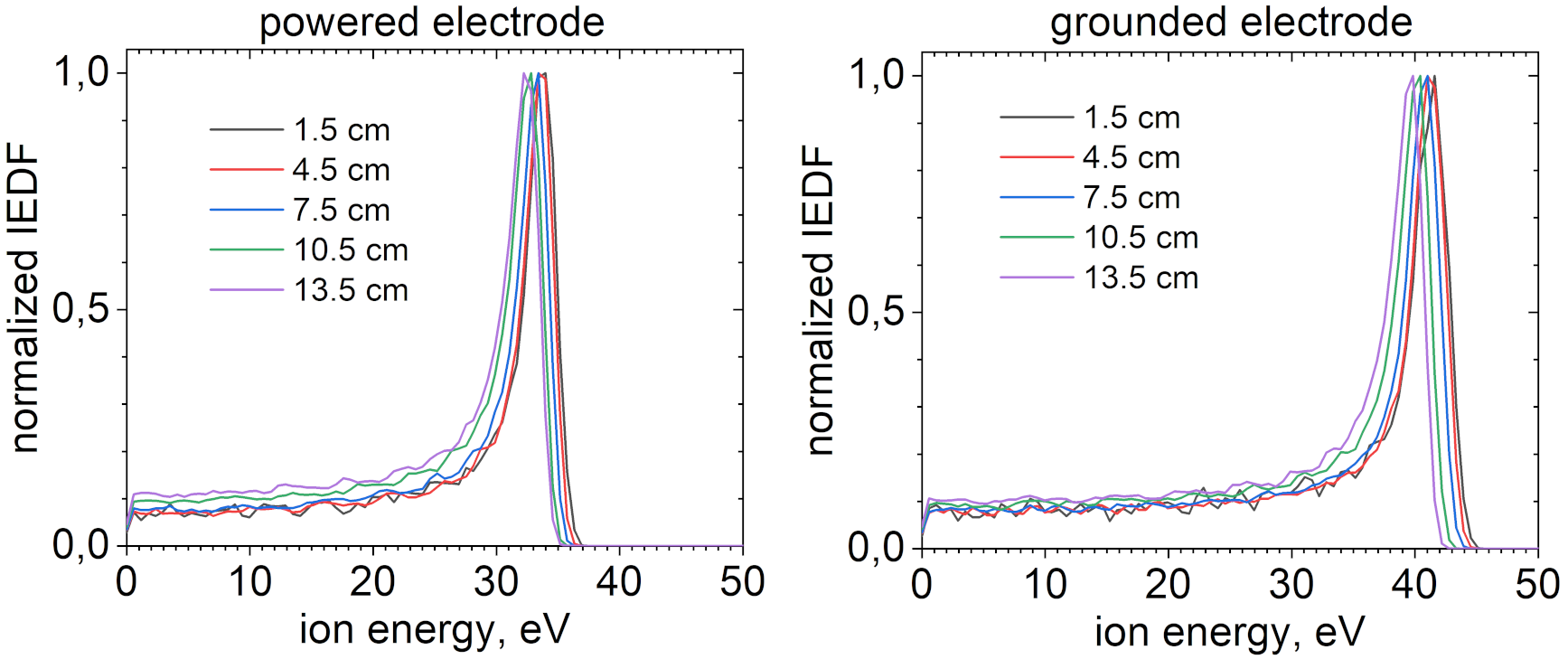}
\caption{IEDFs at the electrodes for the LP $40$ mTorr validation case.}
\label{Fig_IEDF}
\end{figure}

\begin{figure}[h]
\centering
\includegraphics[width=12cm]{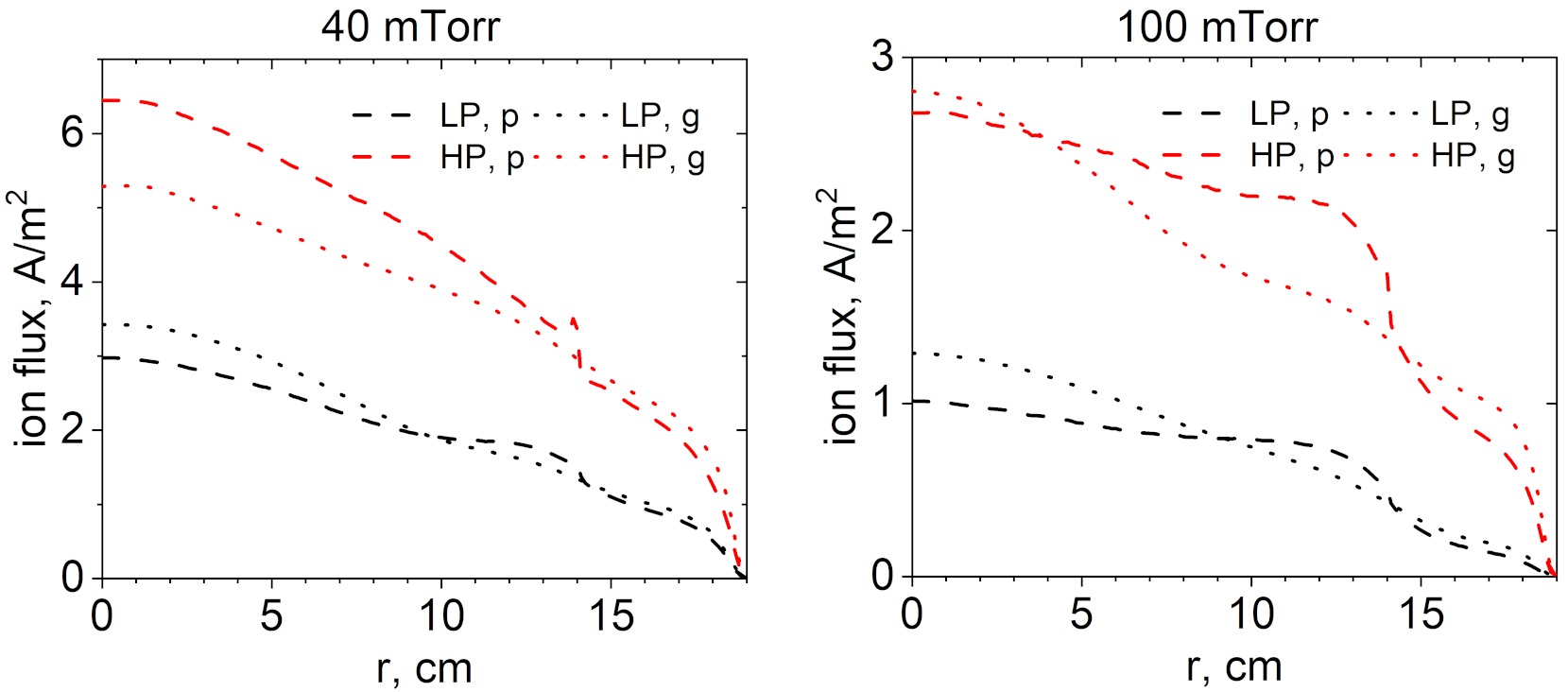}
\caption{Ion fluxes at the electrodes for the validation cases.}
\label{Fig_nonuniform_ion_fluxes}
\end{figure}

\begin{figure}[h]
\centering
\includegraphics[width=6cm]{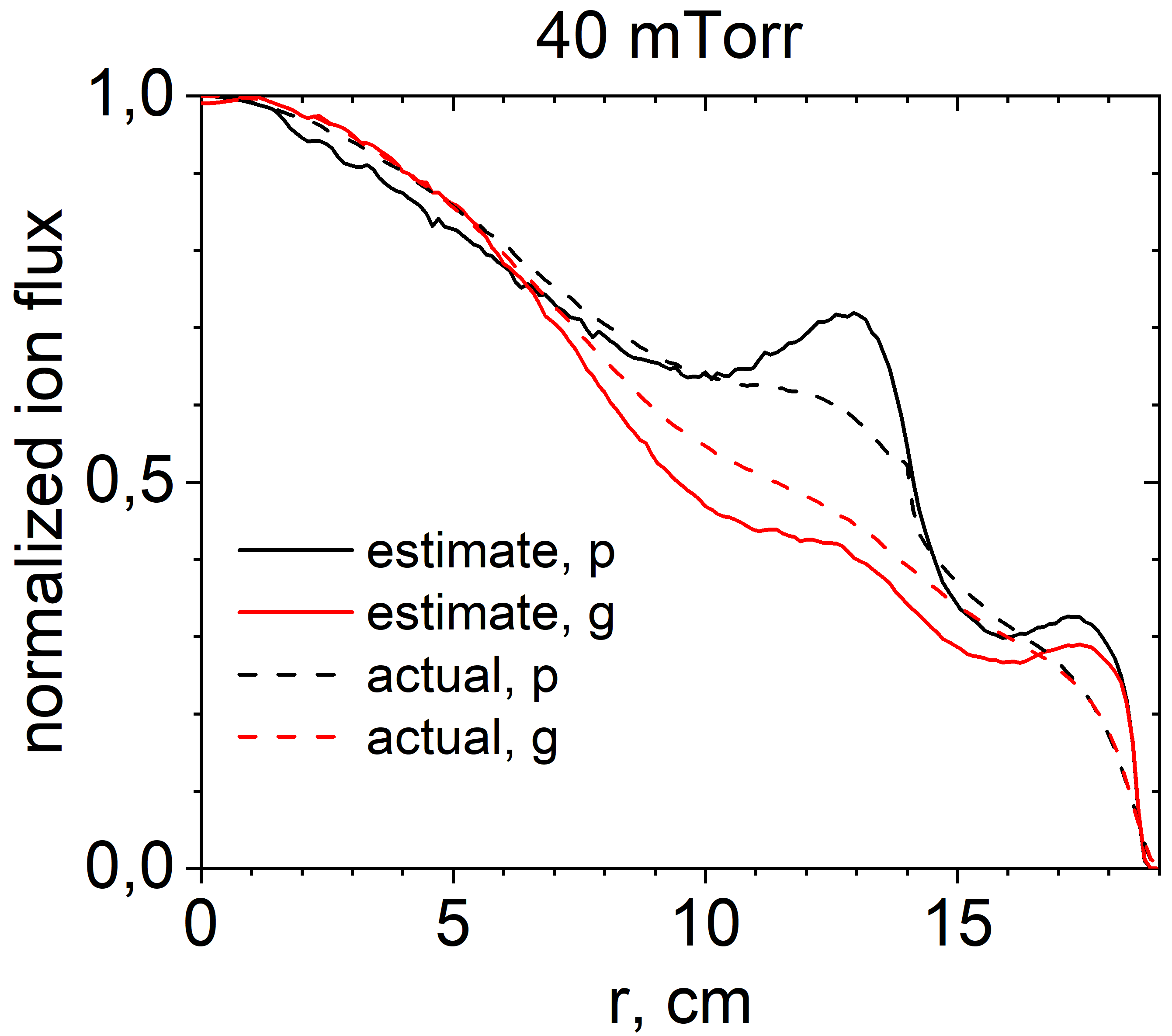}
\caption{Ion flux radial profiles estimated from Eq.(\ref{Eq4_1}).}
\label{Fig_ion_flux_from_ionization_source}
\end{figure}

The nonuniform plasma density observed in the experiments and simulations does not, strictly speaking, have an immediate relevance for the plasma processing technologies. In this respect, the flux and energy of ions emanating from plasma and hitting the reactor electrodes covered with a substrate (and/or a target in case of the sputtering) are more important as they directly affect the quality of plasma processing.    
As one can see from Fig.~\ref{Fig_IEDF} obtained from the simulation of the LP 40 mTorr case, the ion bombardment energy shows peaks that hardly change in the radial direction at both electrodes. A similar observation was made in the experiment investigated in \cite{perret_2005}. It can be explained by an essentially radially uniform dc component of the plasma potential and the equipotentiality of electrodes as long as they are conducting. 
However, it was pointed out in \cite{howling_2005a} that, as long as an electrode is covered with a nonconducting dielectric substrate, this is no longer valid, and the need to locally compensate the total plasma current flowing to the substrate leads to a radially nonuniform ion bombardment energy distribution. In contrast, ion fluxes at the electrodes provided by simulation of the LP 40 mTorr case exhibit a salient radial nonuniformity, see Fig.~\ref{Fig_nonuniform_ion_fluxes}. Note, however, that the corresponding radial profiles are different at different electrodes and change their behavior depending on the neutral gas pressure and power: whereas ion flux at the grounded electrode is typically larger compared to that at the powered electrode for the lower-power cases, the latter can exceed the former if power is increased. Note that at the higher neutral gas pressure of $100$ mTorr the ion flux is noticeably more radially uniform at the powered electrode compared to the grounded electrode. One can link the ion flux radial profiles to the nonuniformity of the ionization source by considering the particle continuity equation for ions. Assuming that the driving frequency is high enough so that the ion density profile is stationary and that the radial ion transport is negligible compared to the axial one, ion flux at each electrode can be estimated through
\begin{equation}
\Gamma_i(r) = \int\limits^{z_e}_{z_*} S(r,z) dz, \label{Eq4_1}
\end{equation}
where $S$ is the ionization rate, $z_e$ is electrode's axial position ($-l$ for the powered and $l$ for the grounded electrode, respectively), and $z_*$ is the axial position where the net ion axial flux vanishes. Fig.~\ref{Fig_ion_flux_from_ionization_source} demonstrates that such a simple estimate indeed yields an ion flux radial profile in reasonable agreement with the simulation results.  

\begin{figure}[h]
\centering
\includegraphics[width=12cm]{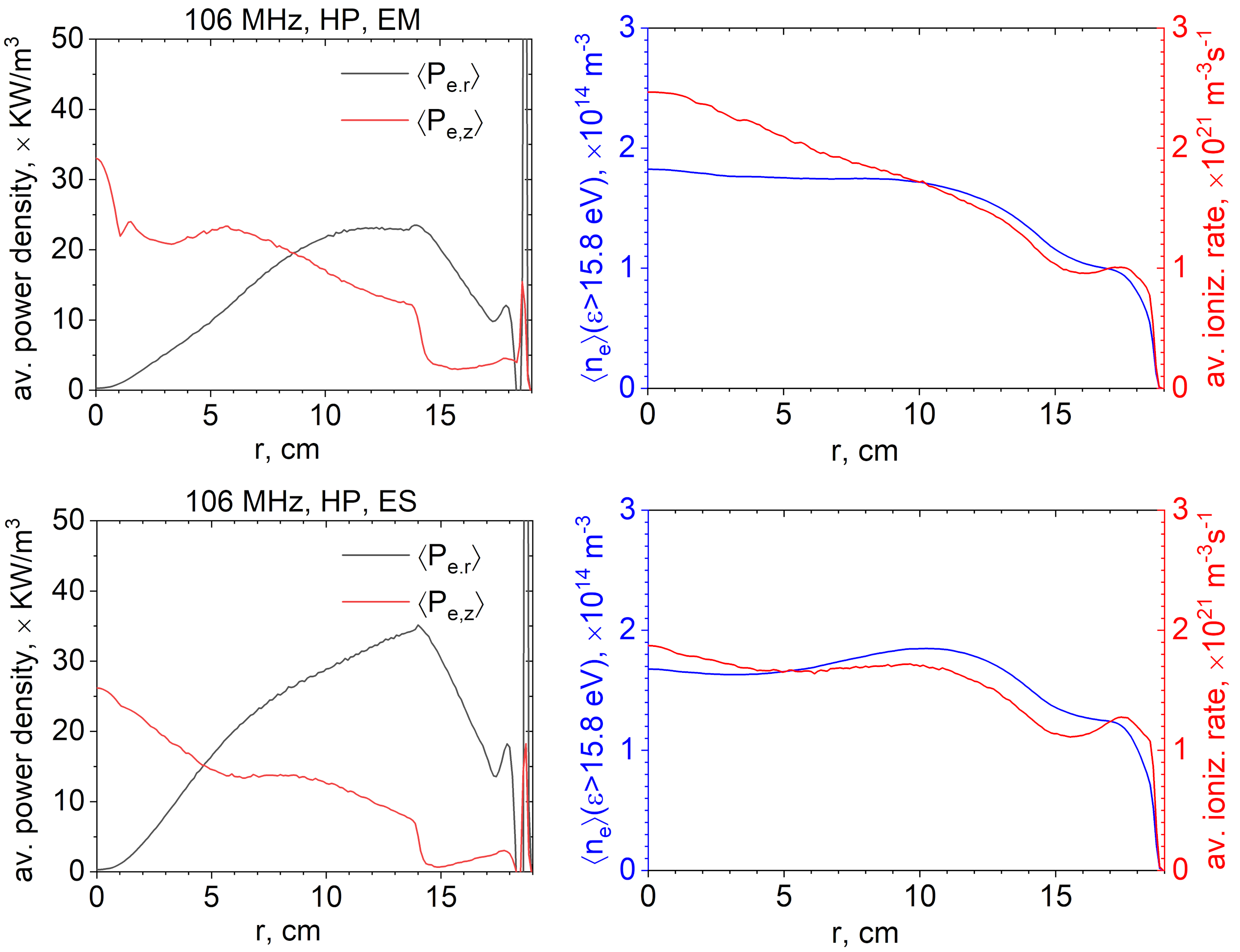}
\caption{Comparison between the radial profiles of the absorbed axial and radial power densities (left column) and between the energetic electron density along
with the ionization rate (right column) obtained from the EM PIC simulation (top row) and the ES PIC simulation (bottom row) for the HP $40$ mTorr case. All corresponding quantities are averaged in the axial direction.}
\label{Fig_avPow_denEnEl_ionizRate_EM_vs_ES}
\end{figure}

\begin{figure}[h]
\centering
\includegraphics[width=12cm]{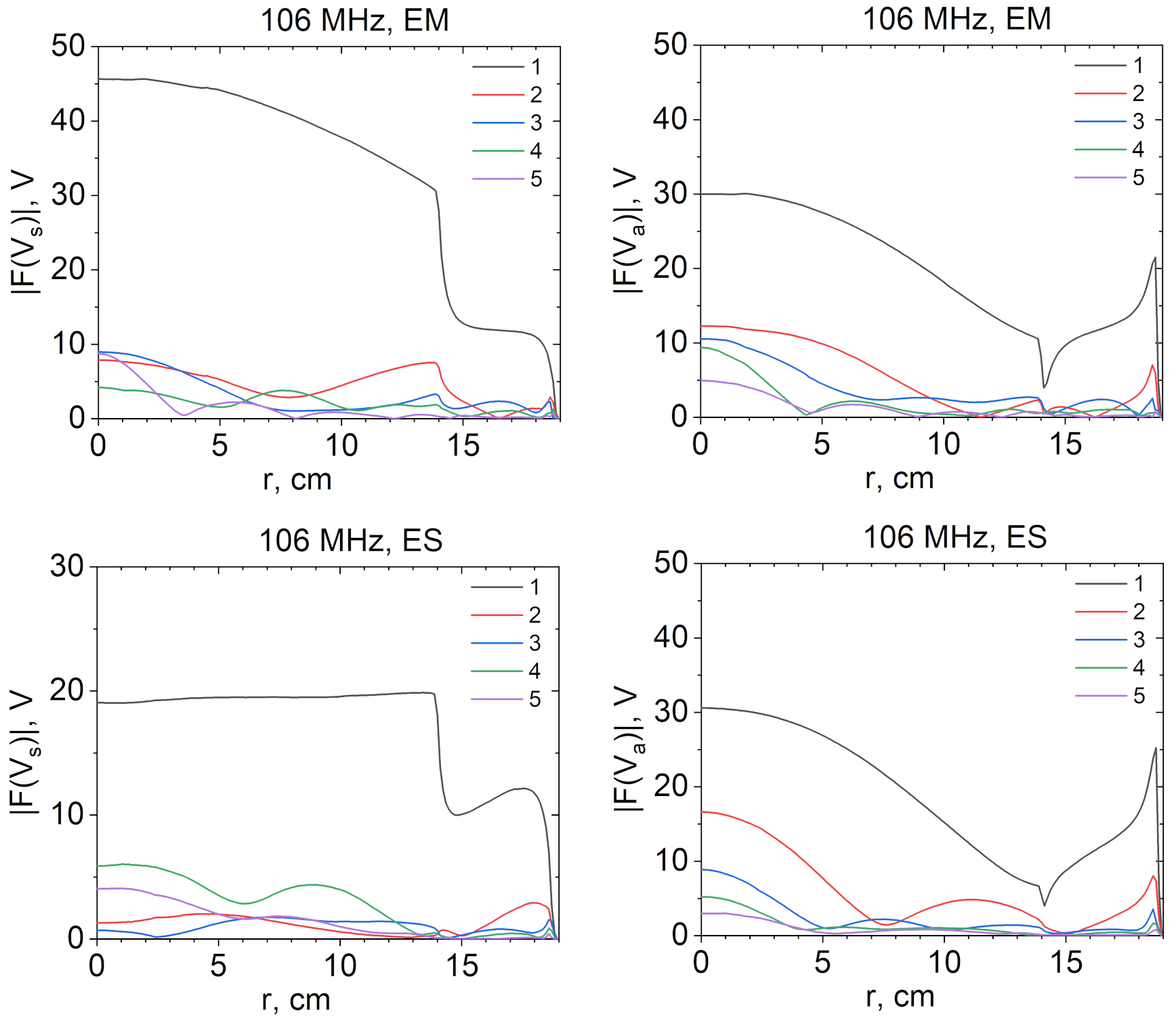}
\caption{Fourier harmonics of the symmetric and asymmetric combinations of the sheath voltages for the simulations shown in Fig.~\ref{Fig_avPow_denEnEl_ionizRate_EM_vs_ES}.}
\label{Fig_Voltage_FB_HP_40mTorr_EM_vs_ES}
\end{figure}

To investigate the factors determining the ionization rate profile, it is instructive to consider the axially averaged power density absorbed by the plasma from the electric field obtained from electromagnetic simulation of the HP case at $40$ mTorr, Fig.~\ref{Fig_avPow_denEnEl_ionizRate_EM_vs_ES}, left, and compare it with the electrostatic counterpart (see also Fig.~\ref{Fig_elDen_HP_40mTorr_EM_to_ES} for the corresponding time-averaged electron density profiles). It can be seen that the electromagnetic and electrostatic models show different averaged power absorption profiles. Power absorption in the radial direction, which for the considered parameters is associated mainly with the asymmetric surface modes \cite{liu_2022}, is greater in the electrostatic case compared to the electromagnetic one. It can be explained by the fact that the asymmetric modes are essentially electrostatic, whereas 
the symmetric mode has a cutoff frequency equal to the plasma series resonance frequency in the electrostatic case, which is absent in the electromagnetic description. This is why the symmetric mode is excited much weaker in the electrostatic case, and the fixed prescribed total power density is matched at the expense of the asymmetric modes, which can also be observed in Fig.~\ref{Fig_Voltage_FB_HP_40mTorr_EM_vs_ES}. Further, Fig.~\ref{Fig_avPow_denEnEl_ionizRate_EM_vs_ES}, right demonstrates that the average density of the energetic electrons having energy above the ionization threshold is relatively similar in terms of the magnitude and the radial profile. It can be attributed to the large mean free path of energetic electrons and similar power absorbed by electrons in both cases. However, the ionization rate radial profile in the electromagnetic case is evidently much more nonuniform than in the electrostatic one and shows a clear center-peaked shape expected from the ``standing wave'' effect \cite{lieberman_2002}. 

\begin{figure}[h]
\centering
\includegraphics[width=7cm]{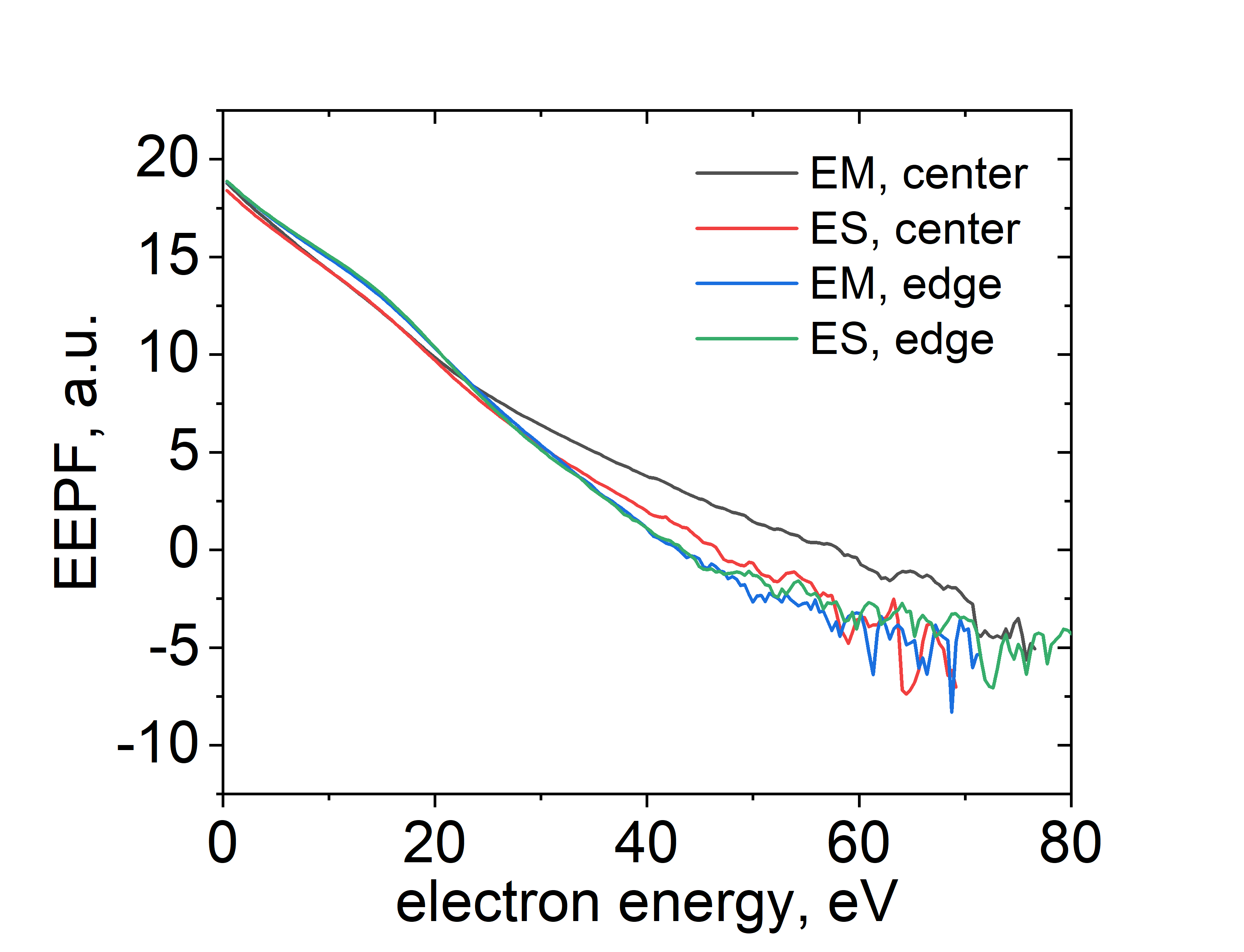}
\caption{EEDFs calculated close to the discharge center and the discharge edge for the simulations shown in Fig.~\ref{Fig_avPow_denEnEl_ionizRate_EM_vs_ES}.}
\label{Fig_EEDF_center_vs_edge_HP_40mTorr}
\end{figure}

%[CHECK!!!] 
This can be explained by the stronger excitation of the symmetric modes in the electromagnetic simulation, the fact that the first symmetric harmonic is much more nonuniform in the electromagnetic case in comparison to the electrostatic one due to the much smaller wavelength, as was argued before (see Fig.~\ref{Fig_Voltage_FB_HP_40mTorr_EM_vs_ES}), and that, as will be argued shortly, the symmetric modes are more efficient in producing the energetic electrons than the asymmetric modes (see Fig.~\ref{Fig_EEDF_center_vs_edge_HP_40mTorr}). 

Fig.~\ref{Fig_Voltage_FB_HP_40mTorr_EM_vs_ES} shows the radial profiles of the Fourier harmonics excited during the electromagnetic and the electrostatic simulations based on the Fourier analysis of the symmetric and asymmetric sheath voltage combinations obtained via $V_s = (V_t+V_b)/2$ and $V_s = (V_t-V_b)/2$ \cite{kawamura_2018,liu_2022}. Each spectrum is dominated by the fundamental harmonic excited at the driving frequency. It can be seen that the asymmetric harmonics are similar in the EM and ES cases, confirming the assumption about the predominantly electrostatic nature of this mode. However, the asymmetric mode in the ES case has a slightly smaller wavelength, which brings it close to the spatially resonant excitation \cite{lieberman_2015}, since the corresponding radial profile is almost continuous at the powered electrode's edge \cite{lieberman_2016,liu_2022}. In this case, the second harmonic seems to have a simultaneous spatial resonance as well, which is also reflected in its relatively large amplitude and results in stronger net radial power absorption in the ES case compared to the EM one seen in Fig.~\ref{Fig_avPow_denEnEl_ionizRate_EM_vs_ES}. In contrast, the symmetric mode harmonics are significantly different between the EM and ES cases. As discussed in Section~\ref{subsec3a}, the symmetric mode dispersion curve in the former case goes to zero in the infinite wavelength limit so that there is an intersection with the driving frequency at a large wavelength, which can cause an enhanced excitation of the corresponding mode if the wavelength is comparable to $\chi_{0,1}/R$.  
The symmetric mode in the ES case for the driving frequency smaller than the cutoff frequency being excited non-resonantly at $k=0$, one can see that amplitude of the symmetric fundamental harmonic is much bigger in the EM compared to the ES case and that the radial profile is flat in the latter case, whereas it falls off to the edge in the former case. Fig.~\ref{Fig_EEDF_center_vs_edge_HP_40mTorr} demonstrates that the EM simulation features a stronger energetic electron tail of the EEDF at the center compared to the ES simulation. It can be linked to the larger amplitude of the fundamental symmetric mode in the EM simulation, which contributes to the more intensive time-averaged axial power absorption at the radial center observed in Fig.~\ref{Fig_avPow_denEnEl_ionizRate_EM_vs_ES}. The symmetric modes, having a strong axial electric field, can efficiently interact with electrons through the sheath motion (Note, however, that the asymmetric modes can also lead to electron acceleration through the same mechanism). The ES case also exhibits higher frequency harmonics featuring multiple nodes suggesting a wave nature. Their origin is not completely clear though, as their frequencies are still smaller than the PSR frequency estimated from the volume-averaged quantities, which yields $2\pi f_{rf}/\omega_{PSR} \approx 0.25$ (which explains the relatively large amplitude of the $4$-th harmonic). It can be speculated that they are driven at the beating frequencies of higher harmonics excited above the PSR frequency. However, a detailed investigation of these modes obtained in the ES limit goes beyond the scope of the present paper. 

\begin{figure}[h]
\centering
\includegraphics[width=12cm]{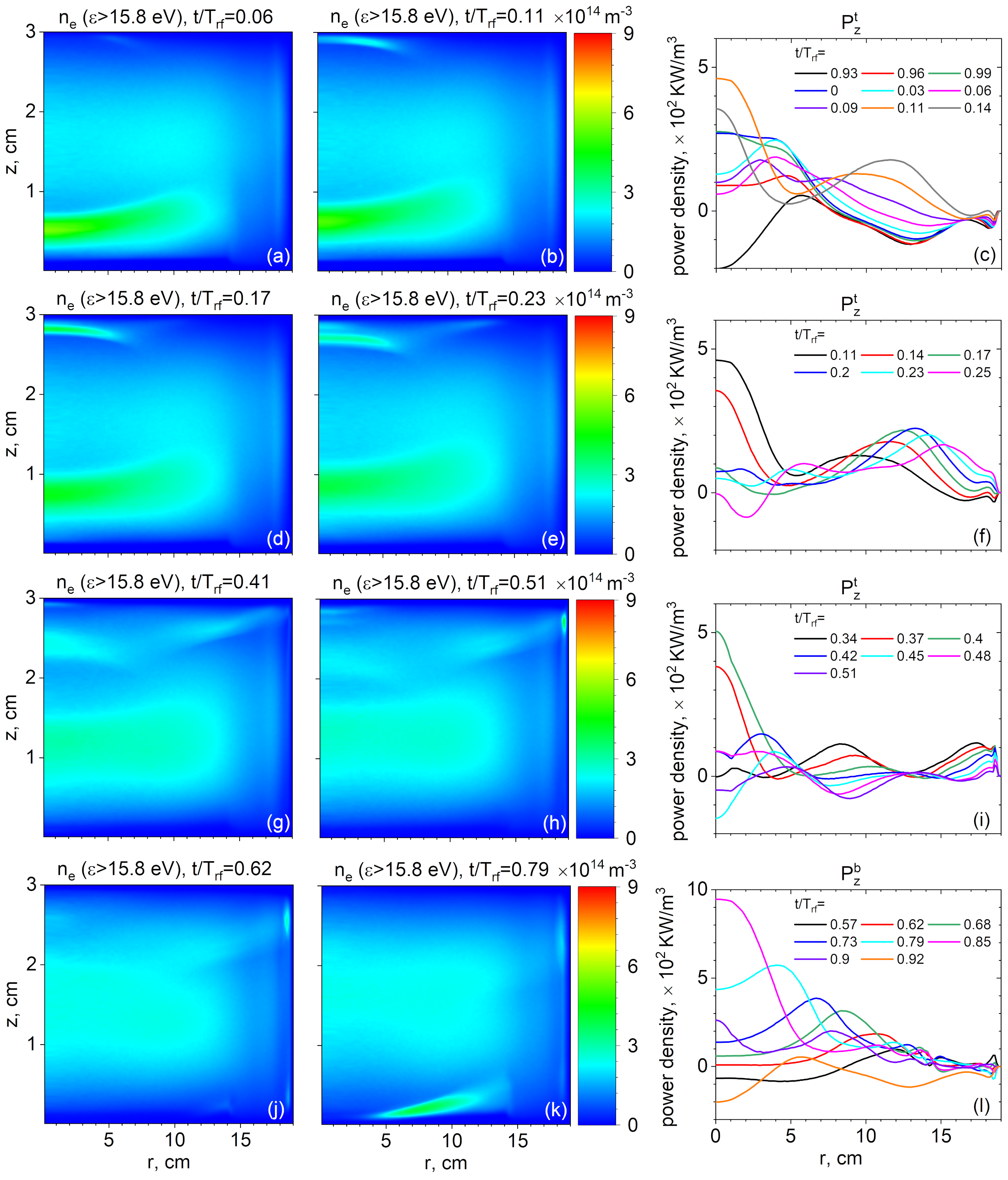}
\caption{Generation of populations of electrons with kinetic energy above the argon ionization threshold as a result of the interaction with the surface modes. The left column shows a snapshot of the energetic electron density when a population begins to form, the middle column shows a similar plot for a moment when the population's formation is well developed, and the right column shows the radial profile of the axially averaged absorbed power density in the axial direction taken at different time moments during the population's formation.}
\label{Fig_denEnEl_and_Pz_vs_time}
\end{figure}

Let us now consider how electrons are energized by the surface modes (by the energization we mean acquiring energy above the ionization threshold \cite{anders_2014,eremin_2021a}). Since the modes can have a strong axial electric field, 
the corresponding mechanism involves interacting with the potential of an expanding or oscillating sheath. It 
is analogous to the efficient electron energization mechanisms well known from the previous literature investigating electron heating in CCPs
in the framework of electrostatic approximation. Of particular interest in this respect is the nonlinear electron resonance heating (NERH)
\cite{mussenbrock_2006,lieberman_2008} and its kinetic interpretation \cite{wilczek_2016} that is based on the interaction of electrons with the sheath oscillations when the PSR is triggered. As the symmetric mode's dynamic in the EM description is a generalization of 
the PSR resonance in the ES approximation, it should also be the case for the corresponding electron heating mechanism. 
Fig.~\ref{Fig_denEnEl_and_Pz_vs_time} shows the generation of energetic electron populations with the kinetic energy above the ionization threshold for argon, which can be ascribed to the interaction of electrons located close to the sheath boundary with the expanding and oscillating sheath, the latter resulting from excitation of different surface modes for the HP $40$ mTorr case simulated with the EM PIC. Four different occurrences of such an interaction were identified. 
The first three rows are related to the electron energization during the grounded sheath expansion, and the fourth row shows a similar process during the powered sheath expansion. For each such events, a corresponding row in Fig.~\ref{Fig_denEnEl_and_Pz_vs_time} shows two time snapshots of the energetic electron density:
One close to the moment when a new energetic population becomes visible on the corresponding numerical diagnostic, and another one when the generation of the corresponding population is in full swing. Besides this, the third plot in each row shows several snapshots of the axial power density averaged over the relevant axial half of the electrode gap, depending on whether the energetic electron population is produced close to the powered or grounded electrode. 

\begin{figure}[h]
\centering
\includegraphics[width=12cm]{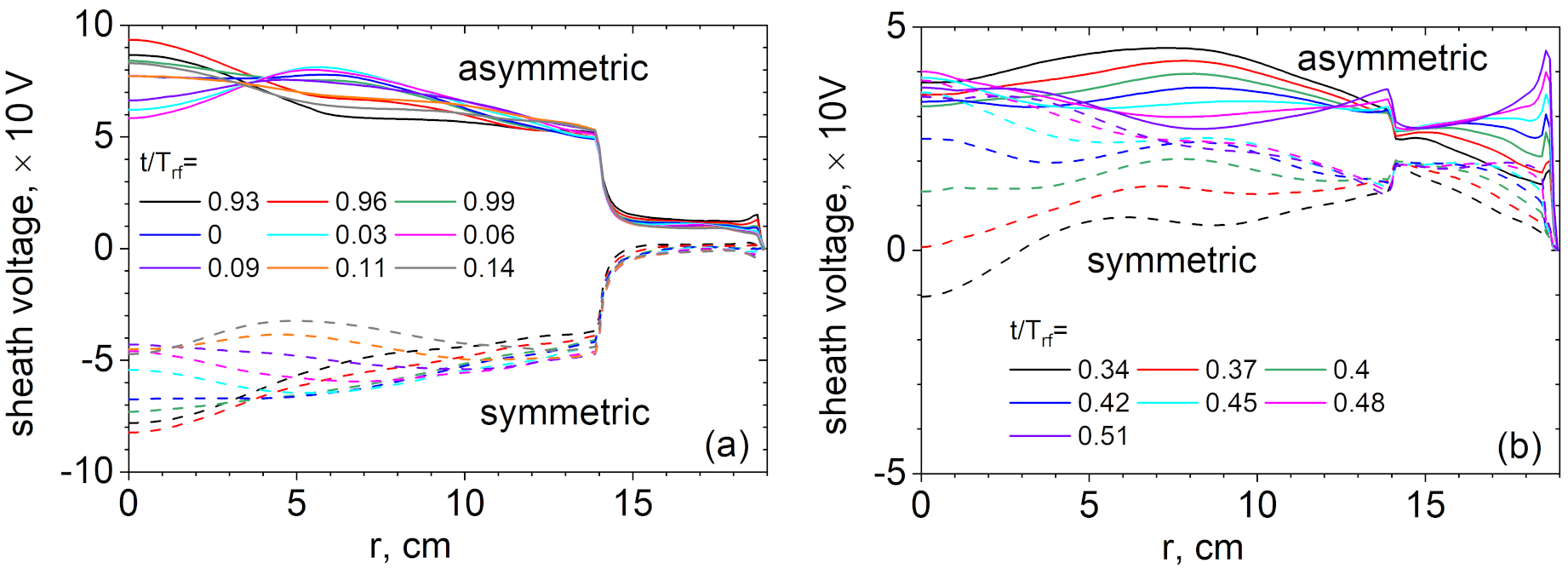}
\caption{Time evolution of the symmetric and asymmetric combinations of the sheath voltages obtained for the cases shown in Fig.~\ref{Fig_denEnEl_and_Pz_vs_time} (c) and (i).}
\label{Fig_mode_identification_sym_and_asym_voltages}
\end{figure}

The electron heating dynamics is evidently rather complicated (compare it with simpler patterns in \cite{eremin_2016} provided by ES PIC/MCC simulations conducted for lower frequencies) and involves a superposition of many excited modes. The first row illustrates a generation of the energetic electron population dominant close to the grounded electrode as a result of the electron acceleration by the grounded sheath undergoing oscillations caused by excitation of a surface wave packet. The oscillatory character of the wave-particle interaction can be seen in Fig.~\ref{Fig_denEnEl_and_Pz_vs_time} (c), which plots the radial profile of the averaged axial heating power density. From this plot, one can see that the time evolution of the corresponding power density profile has almost a standing wave-like pattern with a node around $6$ cm, indicating that the corresponding interaction happens on a time scale smaller than the fundamental frequency (the expanding sheath at the grounded electrode should otherwise lead to a positive power transfer from the electric field to electrons). Note that the corresponding energetic electron population produced as a result of this interaction starts to form from approximately $7$ cm (which correlates with the standing wave's node location) and then propagates towards the center, growing in density as a result of stronger mode oscillation amplitudes closer to the center, which is also reflected in Fig.~\ref{Fig_denEnEl_and_Pz_vs_time} (c). The second row follows the emergence of another energetic electron population generated by a surface mode packet propagating from $7$ cm towards the grounded electrode's edge. Note that Fig.~\ref{Fig_denEnEl_and_Pz_vs_time} (e) shows that the second population is less pronounced than the first and that the first population manifests a secondary structure generated by the oscillating sheath. Fig.~\ref{Fig_denEnEl_and_Pz_vs_time} (f) shows that in this case, the mode packet interacting with 
electrons has a traveling wave character in the region of interest ($\ge 7$ cm). The third row is related to the energization of yet another relatively weak electron group close to the radial center, which takes place some time after the generation of the first two energetic electron populations. Similar to Fig.~\ref{Fig_denEnEl_and_Pz_vs_time} (c), Fig.~\ref{Fig_denEnEl_and_Pz_vs_time} (i) suggests that in this case, the excited mode packet interacting with electrons has a standing wave character. Fig.~\ref{Fig_denEnEl_and_Pz_vs_time} (h) shows that this interaction yields two energetic electron beams propagating towards the plasma bulk. Finally, the fourth row addresses the production of the largest energetic electron population, which occurs this time at the powered electrode. It can be observed that the population starts to emerge at the electrode's edge at $14$ cm, and then the energization process continues towards the
center as the corresponding mode packet propagates in this direction. Based on Fig.~\ref{Fig_denEnEl_and_Pz_vs_time} (l), in this case, the electron-wave interaction seems to be caused by a traveling mode packet, similar to the second row.    

Fig.~\ref{Fig_mode_identification_sym_and_asym_voltages} shows dynamics of the symmetric and asymmetric combinations of the sheath voltages for the first and the third rows of Fig.~\ref{Fig_denEnEl_and_Pz_vs_time} suggestive of surface mode packets in the form of a standing wave leading to the generation of energetic electron groups close to the center. Fig.~\ref{Fig_mode_identification_sym_and_asym_voltages} (a) demonstrates that the dynamics of the asymmetric modes, which form a standing wave, correlate more with the pattern observed in Fig.~\ref{Fig_denEnEl_and_Pz_vs_time} (c) than the symmetric modes forming a traveling wave-like pattern. In contrast to that, Fig.~\ref{Fig_mode_identification_sym_and_asym_voltages} (b) shows that the asymmetric modes seem to build
up a standing wave, but with a node located close to the radial center, whereas the symmetric modes seem to be a result of the fundamental symmetric mode excited at the driving frequency and of a higher-frequency standing wave-like pattern with a node located around $4$ cm, which seems to agree with the pattern observed in Fig.~\ref{Fig_denEnEl_and_Pz_vs_time} (i). These examples conclude that it could be possible to identify the mode type leading to the predominant electron energization: It is of the asymmetric type in the first case, and of the symmetric one in the second. Contrary to \cite{liu_2022}, we do not see any particular reason to unequivocally conclude that the heating/energization of electrons in the axial direction must be dominated by the symmetric modes only. Based on the observations made above, one can see that at a low pressure electron energization is governed by the interaction of electrons with a plasma sheath close to the electrode where it currently expands and with the sheath oscillations when a mode or several modes of any symmetry type are excited during the sheath expansion, which leads to the emergence of energetic electron beams.  
Electron energization in the radial direction is weaker because it mainly occurs through the collision-dominated Ohmic electron heating, whereas 
in the axial direction electrons are energized due to the ``kicks'' imparted by the expanding sheath \cite{popov_1985}, which is an essentially collisionless
process (although collisions might enhance it under some circumstances \cite{schulze_2015}), but requires some collisions to dissipate the energy of electron beams' directed motion \cite{kaganovich_1996,lafleur_2014,lafleur_2015}
(see also recent works on collisionless electron energization in magnetized rf plasmas \cite{eremin_2021a,eremin_2021b,berger_2021}). 

\begin{figure}[h]
\centering
\includegraphics[width=12cm]{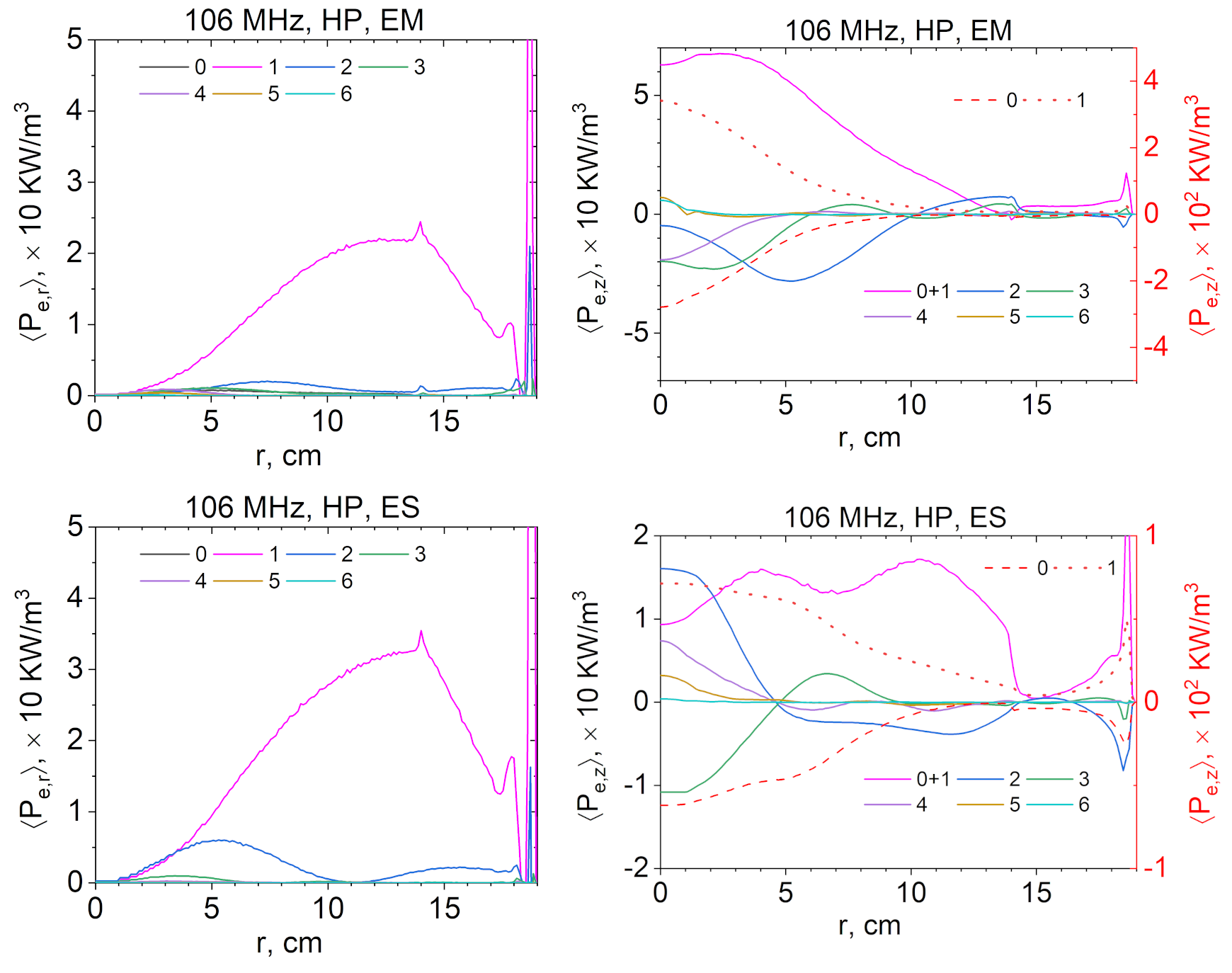}
\caption{Axially and temporally averaged power density profiles for the HP $40$ mTorr case simulated with the EM and ES PIC.}
\label{Fig_av_pow_harmonics}
\end{figure}

It is interesting to see which harmonics significantly contribute to the time-averaged power density absorbed by electrons, see Fig.~\ref{Fig_av_pow_harmonics}. It plots the net (rf period-averaged) power balance of the energy gained by electrons from the electric fields during the phases when electrons are accelerated (e.g., via interaction with the expanding sheaths) and during the phases when electrons return energy to the electric field (e.g., via interaction with the retracting sheaths or due to the decelerating action of the ambipolar field \cite{schulze_2015}).  
This figure demonstrates that in the radial power absorption, the fundamental harmonic makes by far the dominant contribution (note though that the $2$nd harmonic makes a noticeable contribution in the ES case), whereas for the axial power absorption, higher frequency harmonics also play a significant role in shaping the power absorption's radial profile. In the case of the EM simulation, Fig.~\ref{Fig_av_pow_harmonics} shows that the positive power absorption at the fundamental frequency exceeds contributions from all higher harmonics by order of magnitude, but is close in magnitude to the negative
contribution from the zeroth harmonic. The latter can be attributed to the cooling by the ambipolar field at the maximal sheath width boundary when the energetic electron beam arrives there after traveling from the opposite sheath \cite{schulze_2015}). It can also be seen that the only other positive contributions visible close to the center come from the $5$-th and the $6$-th harmonics and are small, although they become noticeable on the net power balance (see Fig.~\ref{Fig_avPow_denEnEl_ionizRate_EM_vs_ES}) when negative contributions from the higher-frequency $2, 3,$ and $4$-th harmonics are added.
Similar trends persist in the radial profile of the axially averaged axial power density profile obtained from the ES simulation. However, 
the net dominant positive contribution from the zeroth and the first harmonics has a much flatter shape, which can be explained by the flat profile of the symmetric mode, which dominates the positive contribution at the fundamental frequency. In this case, another significant positive contribution comes from the $2$-nd harmonic and weaker ones from the $4$-th and the $5$-th harmonics, which demonstrates that the higher harmonics are important in the ES case as well.

\begin{figure}[h]
\centering
\includegraphics[width=15.5cm]{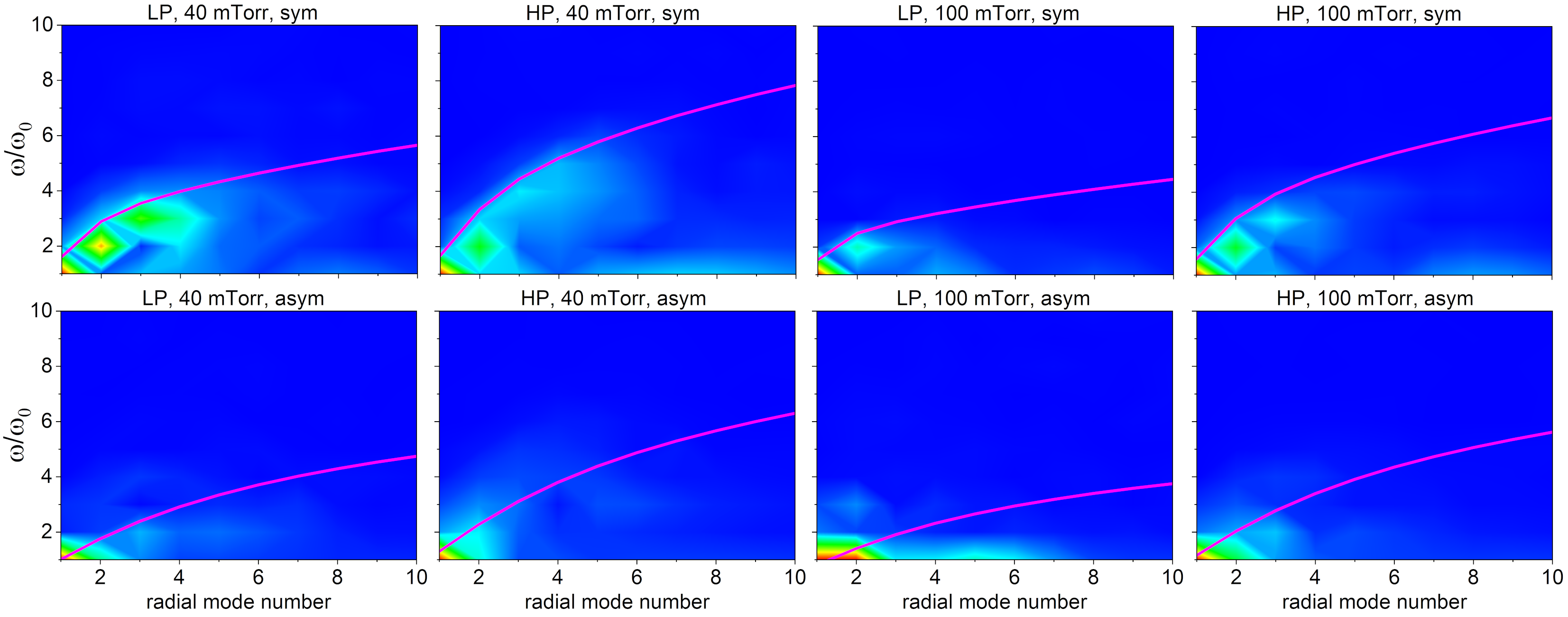}
\caption{Amplitudes of the harmonics obtained after the Fourier decomposition in the temporal and the Fourier-Bessel decomposition in the radial direction,
calculated for the axial current density sampled at the reactor mid-plane for the validation cases shown in
Fig.~\ref{Fig_ne_in_validation_cases}.}
\label{Fig_exc_spectrum_of_jz}
\end{figure}

By performing the Fourier transform in the temporal, and the Fourier-Bessel decomposition in the radial domain (see \ref{Appendix_Fourier_Bessel}), the latter utilizing $\alpha=0$ for the axial, and $\alpha=1$ for the radial component of the conduction current density obtained from the EM PIC simulations of the cases shown in Fig.~\ref{Fig_ne_in_validation_cases} at the reactor mid-plane, it can be traced how the dominant excited harmonics change with the power and the neutral gas pressure, see Fig.~\ref{Fig_exc_spectrum_of_jz}. It can be seen that the higher the collisionality and the lower the power, the fewer harmonics are excited, which is to be expected. Although the analytical expressions were derived for uniform plasma properties, the excited harmonics follow reasonably well the analytical dispersion curves calculated for the unequal sheath widths (derived in \cite{lieberman_2016} and given in \ref{Appendix_dispersion_relation_unequal_sheaths}) with the volume-averaged plasma characteristics estimated from the simulations. One can see that the resonant or near-resonant excitation takes place when the harmonic's frequency is close to the corresponding dispersion curve value at a given radial number. These observations can be explained by the fact that an increased collisionality tends to suppress the surface modes \cite{eremin_2017a} and that a larger power at a fixed driving frequency leads to a larger amplitude of the sheath oscillations, which triggers more surface modes.  

\subsection{Frequency variation of the LP case at $40$ mTorr}

\begin{figure}[h]
\centering
\includegraphics[width=12cm]{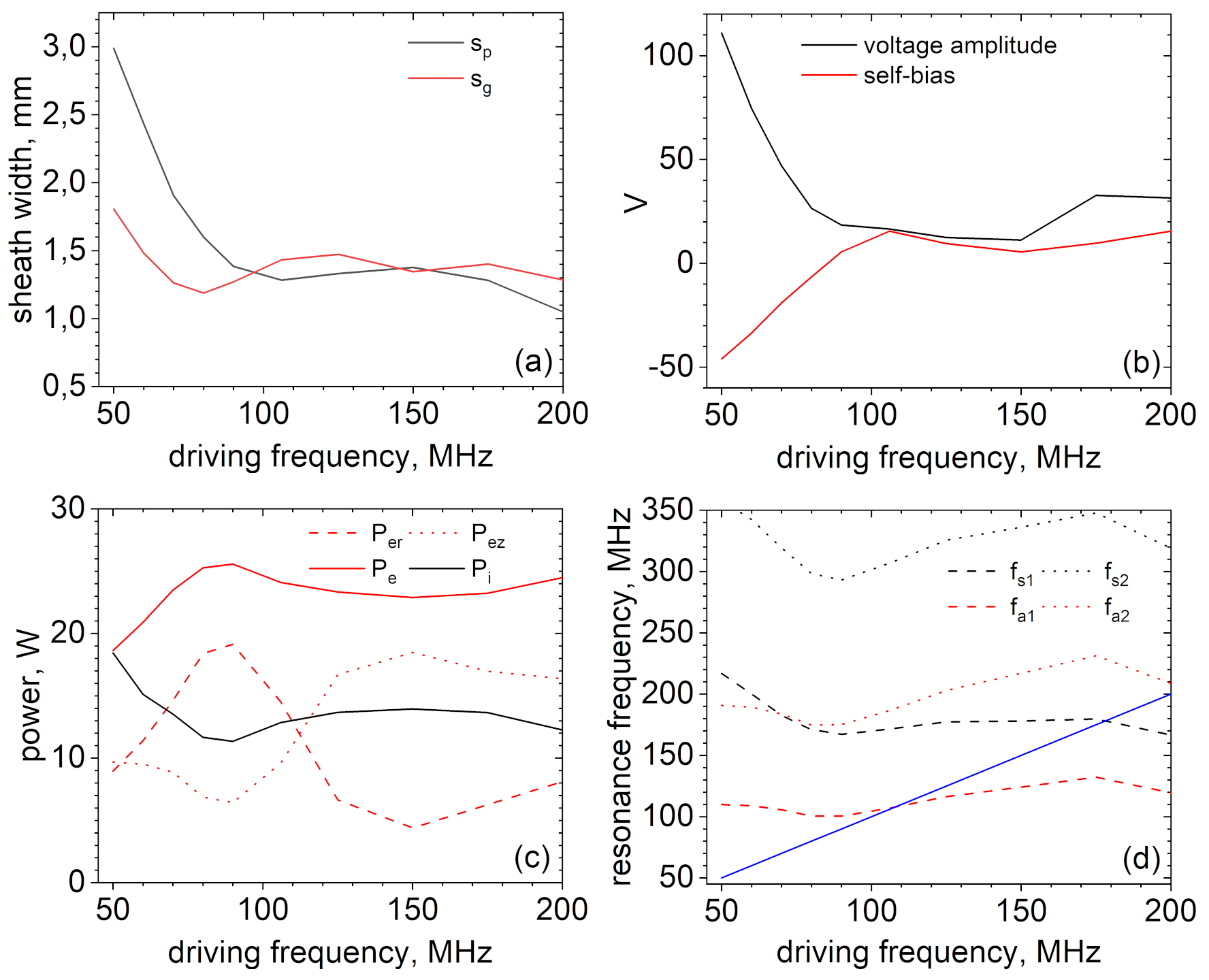}
\caption{Variation of discharge parameters with the driving frequency for the LP $40$ mTorr case: (a) sheath widths, (b) voltage amplitude measured in the reactor chamber at the dielectric spacer between the powered and the grounded electrode, (c) total power absorbed by different plasma species and different directions for the electrons, and (d) resonant frequencies for the first two modes of each type of symmetry calculated from the formulas of \ref{Appendix_dispersion_relation_unequal_sheaths}.}
\label{Fig_Freq_var_plasma_properties}
\end{figure}

One can better understand phenomena taking place in the validation cases by considering a broader range of discharge parameters in simulations, e.g., by considering a frequency variation. In doing this, we focus on the LP $40$ mTorr case. Fig.~\ref{Fig_Freq_var_plasma_properties} plots variation of different plasma parameters obtained from EM PIC simulations in the driving frequency range from $50$ to $200$ MHz. The general trends resemble those obtained recently in \cite{liu_2022} with a different model. Fig.~\ref{Fig_Freq_var_plasma_properties} (a) demonstrates that at low frequencies, the powered sheath is significantly larger than that at the grounded electrode, which is related to the capacitive division of the driving voltage due to the disparate areas of the driven and the grounded electrodes. This is also reflected in the large negative self-bias observed in Fig.~\ref{Fig_Freq_var_plasma_properties} (b). Starting from approx. $100$ MHz, the sheath widths become virtually equal, and the self-bias becomes small but positive. The voltage amplitude needed to match the prescribed power decreases with the driving frequency, featuring a wide and almost flat valley from $90$ to $150$ MHz. The decrease in voltage with frequency is caused by the increased
power absorption efficiency, and it reaches its maximum in the valley, where one or several surface modes are resonant or nearly resonant, which correlates with (c). Fig.~\ref{Fig_Freq_var_plasma_properties} (c) shows the frequency variation of the absorbed power. It can be observed that the power going into the acceleration of ions decreases with frequency, which can be linked to the decreasing sheath voltage, and the power absorbed in electrons goes up. The total power absorbed by electrons can be further split into the radial and axial contributions, which exhibit non-monotonous variations and can be related to various resonances. Indeed, it can be clearly seen that the radial component of the electron absorbed power peaks at $90$ MHz while the axial component has a minimum there, and the components swap these behaviors at $150$ MHz. Following \cite{liu_2022}, one can explain this by the fundamental resonance with the first asymmetric harmonic at $90$ MHz and with the first symmetric harmonic around $150$ MHz. Whereas we have argued previously that not only symmetric modes can significantly contribute to the axial absorption, the case with $150$ MHz is close to the symmetric mode resonance, and close to this frequency the axial power absorption is dominated by the first symmetric harmonic. This is corroborated by Fig.~\ref{Fig_Freq_var_plasma_properties} (d) depicting the resonant frequencies for the first two symmetric and asymmetric harmonics calculated from the formulas of \ref{Appendix_dispersion_relation_unequal_sheaths} with plasma parameters obtained from the corresponding EM PIC simulations. Whenever the resonant frequencies coincide with the driving frequency, one has a resonance for that mode. These points are located at the intersection of the straight blue line and the curves. One can see that this analysis predicts resonant excitation of the first asymmetric harmonic around $100$ MHz, and of the first symmetric harmonic around $175$ MHz. If the power is increased so that higher frequency harmonics can be excited by the sheath motion, strong resonances are also possible at higher harmonics of the fundamental frequency.
%can occur for smaller driving frequencies or with a surface mode having a higher radial mode number (the resonant curves, however, would need to be recalculated from the corresponding new simulations). 

\begin{figure}[h]
\centering
\includegraphics[width=12cm]{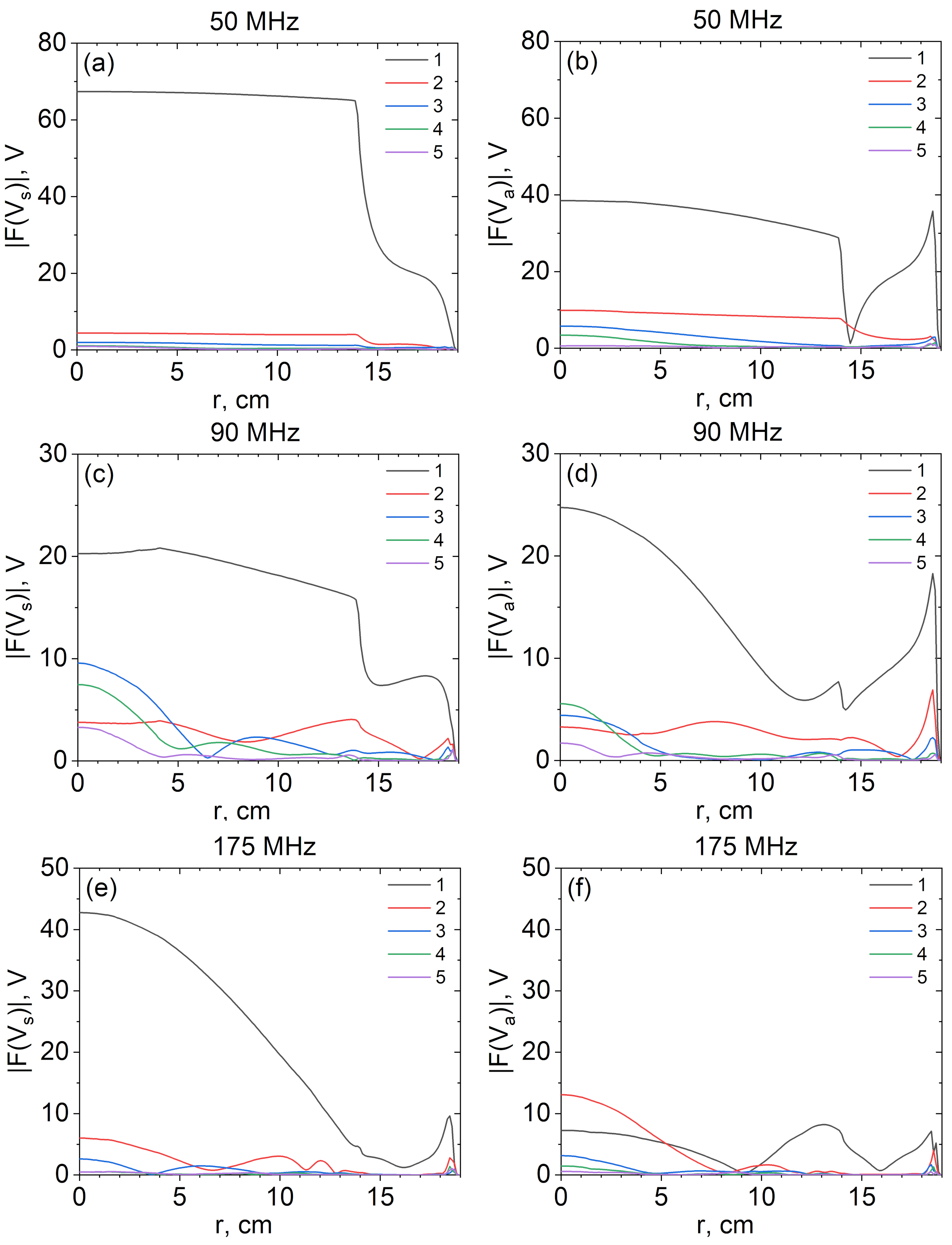}
\caption{Fourier harmonics of the symmetric and asymmetric combinations of the sheath voltages for a number of LP $40$ mTorr cases with various frequencies: non-resonant case at $50$ MHz (top row), first asymmetric resonance at $90$ MHz (middle row), first symmetric resonance at $175$ MHz (bottom row).}
\label{Fig_Fourier_harmonics_sym_and_asym_sheath_voltage}
\end{figure}

Finally, one can see where the spatial resonances occur at the fundamental frequency for the asymmetric and symmetric modes by looking at the 
Fourier harmonics of the asymmetric and symmetric sheath voltage combinations. To back up this point, Fig.~
\ref{Fig_Fourier_harmonics_sym_and_asym_sheath_voltage} shows the radial profiles of the corresponding quantities for three selected driving frequencies, 
a non-resonant low-frequency case at $50$ MHz, the case with the near-resonant asymmetric fundamental harmonic at $90$ MHz, and the case with the
resonant symmetric fundamental harmonic at $175$ MHz. One can judge whether a harmonic is resonant or not by estimating the discontinuity at the
powered electrode's edge at $14$ cm. All plots confirm the expectations. In addition to this, the high-frequency case with $175$ MHz exhibits a very weak excitation of the asymmetric mode, so that the assumption that close to that frequency the power absorption is dominated by the symmetric modes is valid.

%Following previous literature (e.g., \cite{liu_2022}), this can be explained by an increased power absorption efficiency due to different modes, which are resonant or nearly resonant in this range of frequencies    

%%%%%%%%%%%%%%%%%%%%%%
%\clearpage
\section{Conclusions} \label{sec5}

This work presents for the first time results of modeling the capacitively coupled discharges with large electrode areas and driven with very high frequencies at a low pressure employing a fully electromagnetic particle-in-cell code. The code used is based on the energy- and charge-conserving implicit method and is verified with three different tests checking its consistency in reproducing known effects. In particular, the considered effects include the analytical dispersion curves of surface modes excited
in such discharges and affecting plasma uniformity. The code is further validated using the experimental data obtained earlier in \cite{sawada_2014}, showing
a good agreement with the experiment. It is argued that the particle-in-cell approach is a high-fidelity method, which has many advantages and captures effects important for modeling the plasmas in question. It is possible due to a self-consistent kinetic and nonlocal description and taking into account electron inertia. 

The simulated cases are explored further using additional diagnostics. Special attention is given to the uniformity of plasma parameters having immediate
importance for the plasma processing technologies, such as the energy distributions and fluxes of ions impinging on the electrodes. It is demonstrated that for conducting electrodes considered in the present work, the energy distribution of ions hitting the electrodes peaks at approximately the same energy at all radial locations on the electrode surface. In contrast, the ion fluxes to the electrodes are pronouncedly nonuniform in the radial direction. The corresponding profile is shaped mainly by the ionization source, which in turn is governed by the electron energization dynamics. The electrons are energized due to the interaction with the surface modes of two axial electric field symmetry types. The asymmetric modes energize electrons predominantly via the collision-dominated Ohmic heating caused by the radial electric field. The symmetric modes accelerate electrons mainly through an essentially collisionless mechanism related to the interaction of electrons with a moving plasma sheath, which efficiently produces beams of energetic electrons propagating toward the plasma bulk, clearly seen in the simulations. It is observed that the asymmetric modes can also contribute to this mechanism.
The numerical data showed that at the considered frequencies, electrons are mainly accelerated by expanding sheaths which is related to the discharge dynamics at the fundamental driving frequency. However, modes driven at higher frequency harmonics or at larger radial mode numbers can also contribute to electron acceleration by causing the sheath oscillations whenever the corresponding modes are triggered. 
%The modes of both symmetry types can participate in such processes and 
It is demonstrated that the discharge exhibits a broad spectrum of excited frequency harmonics for the considered parameters. A smaller power and a larger pressure appear to decrease the number of excited harmonics due to a smaller amplitude of the sheath oscillations and the suppression of high frequency harmonics due to collisions. The electron energization in the axial direction leads to the generation of electrons with greater energies compared to the electron energization occurring in the radial direction due to the higher efficiency of the sheath-based energization mechanism in the axial direction compared to the Ohmic heating mechanism acting in the radial direction.    

Finally, a frequency variation for a selected case demonstrates that in the considered range of parameters, the excitation of surface modes approximately follows the
analytical predictions obtained in the previous literature and that the time-averaged power absorbed by electrons exhibits peaks of the radial component whenever an asymmetric mode has a spatial resonance or of the axial component whenever a spatial resonance of a symmetric mode is expected. 

The observed plasma dynamics and the wave-particle interaction patterns are rather complicated and involve the excitation of many surface modes, which makes the electromagnetic particle-in-cell method, which is self-consistent, kinetic, and nonlocal, a promising tool for predictive modeling.

%\begin{itemize}
%\item Modes are excited dynamically and have form of travelling waves, standing wave patterns are not observed.
%\item Dominant modes are excited at different locations (different sheaths and either at the center or at the periphery) depending on different power.
%\item Mode excitation is triggered by the sheath motion, the sheath edge exhibits oscillatory or an abrupt change in behavior (similar to \cite{wilczek_2016}). 
%\item Ion flux radial uniformity is different at different electrodes. 
%\item Observation of the parallel resonance excitation is only possible for a PIC model and absent from the electron drift-diffusion-based models.
%\end{itemize}

%\begin{equation}
%	\cases{
%		 \Delta y < \frac{\lambda_D}{2}, \\
%		 \Delta t < \frac{0.2}{\omega_{pe}},
%	}
%	\label{Eq_PIC_constraints}
%\end{equation}

%\begin{figure}[!htb]
%	\centering
%	\includegraphics{2D_numerical_setup}
%	\caption{2D radial-azimuthal ($z,y$) setup.}
%	\label{Fig_L2B_2D_numerical_setup}	
%\end{figure} 

\section*{Acknowledgments}

The authors gratefully acknowledges support by DFG (German Research Foundation) within the SFB-TR 87 project framework and by Tokyo Electron Technology Solutions Limited.  

\appendix 

\section{Fourier-Bessel series} \label{Appendix_Fourier_Bessel}

Due to the peculiarities of cylindrical geometry, the spatially sinusoidal Fourier harmonics relevant for rectangular geometry \cite{eremin_2017a,eremin_2017b} are replaced by functions comprising the Fourier-Bessel series \cite{Schroeder_1993}, $F_m^\alpha(r) = J_\alpha(\chi_{\alpha,m}r/R)$ with $J_\alpha$ being the corresponding Bessel function and $\chi_{\alpha,m}$ its $m$-th zero. It can be expected that if a harmonic has a wavelength lying close to the lateral reactor length, then such a harmonic would be resonantly excited and will have a large amplitude \cite{lieberman_2015,eremin_2017a,eremin_2017b}. In cylindrical geometry such a resonance condition leads to a condition for the ``radial wavenumber'' $k^s_{r,m} = \chi_{0,m}/R$ for the symmetric mode and to $k^a_{r,m} = \chi_{1,m}/R$ for the antisymmetric mode \cite{lieberman_2015,lieberman_2016,kawamura_2018,liu_2021a}. However, one needs to note that the radial dependence, for example, of the electric field components $E_r$ and $E_z$, is described by the Bessel functions in cylindrical geometry, so that the meaning of the wavenumber becomes somewhat loosely defined in comparison to rectangular geometry. Drawing an analogy with rectangular geometry, it appears meaningful to represent various quantities having a radial dependence in terms of the Fourier-Bessel series,
\begin{equation}
f(r) \sim \sum\limits_{m=1}^{m_{\rm max}} c_m F_m^\alpha(r), \label{AppA_eqA1}
\end{equation}
where coefficients $c_m$ are to be calculated from
\begin{equation}
c_m = \frac{\int\limits^R_0 dr r f(r) F_m^\alpha(r)}{\int\limits^R_0 dr r \left(F_m^\alpha(r)\right)^2}, \label{AppA_eqA2}
\end{equation}
and $\alpha$ in $F_m$ should be chosen based on the expected behavior of the quantity (e.g., $\alpha=0$ for $E_z$ and $\alpha=1$ for $E_r$).
Such a decomposition is arguably not complete, because it results in a function which goes to zero at $r=R$, whereas the underlying function $f$
in general does not have to exhibit such a behavior (e.g., $E_r$). Another situation when the decomposition does not work well is when the function to be decomposed is relatively flat and abruptly goes to zero at the edge (e.g., $E_z$ at low driving frequencies).
In this case one can see that the decomposition works poorly whenever a wide spectrum is observed. However, it was found that in practice it rarely occurs and that typically only a few first Fourier-Bessel resonantly excited harmonics dominate the spectrum. It also proved convenient to characterize the spectral harmonics by their corresponding ``radial mode number'' $m$. Note, however, that the corresponding radial harmonics $F_m^\alpha$ might differ depending on the choice of $\alpha$.

\section{Dispersion relation of the surface modes when sheath widths are not equal} \label{Appendix_dispersion_relation_unequal_sheaths}

Eqs.(\ref{eq3_1}) governing the dispersion relation of the propagating symmetric and asymmetric surface modes need to be modified  if unequal sheath widths are assumed for the powered and the grounded sheath, respectively \cite{lieberman_2016}. Denoting the former as $s$ and the latter as $w$ and eliminating the $\delta$ parameter from Eqs.(14)-(17) in \cite{lieberman_2016}, one obtains generalized dispersion relations for the symmetric and asymmetric modes. In the notations adopted in this paper they read
\begin{equation}
\tanh^{-1}\left(-\frac{\epsilon_p\alpha}{\beta}\tanh(\alpha w)\right)+\tanh^{-1}\left(-\frac{\epsilon_p\alpha}{\beta}\tanh(\alpha s)\right)
= \beta(2l-s-w) \label{AppB_eqB1}
\end{equation}
and 
\begin{equation}
\coth^{-1}\left(-\frac{\epsilon_p\alpha}{\beta}\tanh(\alpha w)\right)+\coth^{-1}\left(-\frac{\epsilon_p\alpha}{\beta}\tanh(\alpha s)\right)
= \beta(2l-s-w), \label{AppB_eqB2}
\end{equation}
respectively. It is easy to see that these equations become Eqs.(\ref{eq3_1}) if $s=w$.

\section*{References}
\bibliographystyle{ieeetr}
%\bibliographystyle{unsrt}
%\begin{thebibliography}{99}

\bibliography{references}
%\end{thebibliography}

\end{document}

%% file: customed_settings.tex
\usepackage[utf8]{inputenc}
\usepackage[T1]{fontenc}

\usepackage{hyperref}
\hypersetup{
    colorlinks=true,
    linkcolor=blue,
    filecolor=magenta,      
    urlcolor=cyan,
    citecolor=blue}

\usepackage{graphicx}
\usepackage[justification=centering]{caption}
\usepackage[justification=centering]{subcaption}
\usepackage{siunitx}
\DeclareSIUnit\gauss{G}
\usepackage{booktabs}
\newcolumntype{C}{>{$}c<{$}}
\newcolumntype{L}{>{$}l<{$}}
\newcolumntype{R}{>{$}r<{$}}
\usepackage[noabbrev]{cleveref}

%% file: EM_PIC_paper.bbl
\begin{thebibliography}{100}

\bibitem{sawada_2014}
I.~Sawada, {\relax P.L.G}.~Ventzek, B.~Lane, T.~Ohshita, {\relax
  R.R}.~Upadhyay, and {\relax L.L}.~Raja, ``Relationship between center-peaked
  plasma density profiles and harmonic electromagnetic waves in very high
  frequency capacitively coupled plasma reactors,'' {\em Jpn. J. Appl. Phys.},
  vol.~53, p.~03DB01, 2014.

\bibitem{lieberman_2005}
{\relax M.A}.~Lieberman and {\relax A.J}.~Lichtenberg, {\em Principles of
  Plasma Discharges and Materials Processing}.
\newblock Wiley, 2nd ed.~ed., 2005.

\bibitem{chabert_2011}
P.~Chabert and N.~Braithwaite, {\em Physics of Radiofrequency Plasmas}.
\newblock Cambridge University Press, 2011.

\bibitem{makabe_2006}
T.~Makabe and {\relax Z.L}.~Petrovic, {\em Plasma Electronics: Applications in
  Microelectronic Device Fabrication}.
\newblock Taylor and Francis Ltd., 2006.

\bibitem{stevens_1996}
J.~Stevens, M.~Sowa, and J.~Cecchi, ``Uniformity of radio frequency bias
  voltages along conducting surfaces in a plasma,'' {\em J. Vac. Sci. Technol.
  A}, vol.~14, no.~1, p.~139, 1996.

\bibitem{schmitt_2002}
J.~Schmitt, M.~Elyaakoubi, and L.~Sansonnens, ``Glow discharge processing in
  the liquid crystal display industry,'' {\em Plasma Sources Sci. Technol.},
  vol.~11, p.~A206, 2002.

\bibitem{perret_2003}
A.~Perret, P.~Chabert, {\relax J.-P}.~Booth, J.~Jolly, J.~Guillon, and
  P.~Auvray, ``Ion flux nonuniformities in large-area high-frequency capacitive
  discharges,'' {\em Appl. Phys. Lett.}, vol.~83, p.~243, 2003.

\bibitem{bowers_2001}
K.~Bowers, ``High frequency electron resonances and surface waves in
  unmagnetized bounded plasmas,'' {\em Ph.D. thesis, UC Berkeley, 2001}.

\bibitem{lieberman_2002}
{\relax M.A}.~Lieberman, {\relax J.P}.~Booth, P.~Chabert, {\relax J.M}.~Rax,
  and {\relax M.M}.~Turner, ``Standing wave and skin effects in large-area,
  high-frequency capacitive discharges,'' {\em Plasma Sources Sci. Technol.},
  vol.~11, p.~283, 2002.

\bibitem{sansonnens_2006}
L.~Sansonnens, {\relax A.A}.~Howling, and {\relax Ch}.~Hollenstein,
  ``Electromagnetic field nonuniformities in large area, high-frequency
  capacitive plasma reactors, including electrode asymmetry effects,'' {\em
  Plasma Sources Sci. Technol.}, vol.~15, p.~302, 2006.

\bibitem{kawamura_2018}
E.~Kawamura, {\relax M.A}.~Lieberman, and {\relax A.J}.~Lichtenberg, ``Symmetry
  breaking in high frequency, symmetric capacitively coupled plasmas,'' {\em
  Phys. Plasmas}, vol.~25, p.~093517, 2018.

\bibitem{dvinin_2021}
S.~Dvinin, O.~Sinkevich, Z.~Kodirzoda, and D.~Solikhov, ``Features of
  electromagnetic field excitation in a capacitive hf discharge. ii. symmetric
  discharge completely filling vacuum chamber under symmetric and asymmetric
  excitation,'' {\em Plasma Physics Reports}, vol.~47, p.~28, 2021.

\bibitem{howling_2004}
{\relax A.A}.~Howling, L.~Sansonnens, J.~Ballutaud, and {\relax
  Ch}.~Hollenstein, ``Nonuniform radio-frequency plasma potential due to edge
  asymmetry in large-area radio-frequency reactors,'' {\em J. Appl. Phys.},
  vol.~96, no.~10, p.~5429, 2004.

\bibitem{lieberman_2016}
M.~Lieberman, A.~Lichtenberg, E.~Kawamura, and P.~Chabert, ``Linear
  electromagnetic excitation of an asymmetric low pressure capacitive discharge
  with unequal sheath widths,'' {\em Phys. Plasmas}, vol.~23, p.~013501, 2016.

\bibitem{howling_2005}
{\relax A.A}.~Howling, L.~Derendinger, L.~Sansonnens, H.~Schmidt, {\relax
  Ch}.~Hollenstein, E.~Sakanaka, and {\relax J.P.M}.~Schmitt, ``Probe
  measurements of plasma potential nonuniformity due to edge asymmetry in
  large-area radio-frequency reactors: The telegraph effect,'' {\em J. Appl.
  Phys.}, vol.~97, p.~123308, 2005.

\bibitem{eremin_2017b}
D.~Eremin, {\relax R.P}.~Brinkmann, and T.~Mussenbrock, ``Observations of
  surface mode influence on plasma uniformity in {PIC/MCC} simulations of large
  capacitive discharges,'' {\em Plasma Process Polym.}, vol.~14, p.~1600164,
  2017.

\bibitem{howling_2007}
L.~Sansonnens, {\relax A.A}.~Howling, and {\relax Ch}.~Hollenstein,
  ``Electromagnetic sources of nonuniformity in large area capacitive
  reactors,'' {\em Thin Solid Films}, vol.~515, p.~5059, 2007.

\bibitem{chabert_2004}
P.~Chabert, J.~Raimbault, J.~Rax, and M.~Lieberman, ``Self-consistent nonlinear
  transmission line model of standing wave effects in a capacitive discharge,''
  {\em Phys. Plasmas}, vol.~11, p.~1775, 2004.

\bibitem{eremin_2016}
D.~Eremin, S.~Bienholz, D.~Szeremley, J.~Trieschmann, S.~Ries, P.~Awakowicz,
  T.~Mussenbrock, and {\relax R.P}.~Brinkmann, ``On the physics of a large ccp
  discharge,'' {\em Plasma Sources Sci. Technol.}, vol.~25, p.~025020, 2016.

\bibitem{eremin_2017a}
D.~Eremin, ``Modeling of resonant surface wave excitation in a large ccp
  reactor,'' {\em IEEE Transactions on Plasma Science}, vol.~45, no.~4, p.~527,
  2017.

\bibitem{wen_2017b}
D.-Q. Wen, E.~Kawamura, M.~Lieberman, A.~Lichtenberg, and Y.-N. Wang,
  ``Two-dimensional particle-in-cell simulations of standing waves and
  wave-induced hysteresis in asymmetric capacitive discharges,'' {\em J. Phys.
  D: Appl. Phys.}, vol.~50, p.~495201, 2017.

\bibitem{chabert_2006}
P.~Chabert, J.~Raimbault, P.~Levif, J.~Rax, and M.~Lieberman, ``Inductive
  heating and e to h transitions in high frequency capacitive discharges,''
  {\em Plasma Sources Sci. Technol.}, vol.~15, p.~S130, 2006.

\bibitem{upadhyay_2013}
R.~Upadhyay, I.~Sawada, P.~Ventzek, and L.~Raja, ``Effect of electromagnetic
  waves and higher harmonics in capacitively coupled plasma phenomena,'' {\em
  J. Phys. D: Appl. Phys.}, vol.~46, p.~472001, 2013.

\bibitem{miller_2006}
P.~Miller, E.~Barnat, G.~Hebner, A.~Paterson, and J.~Holland, ``Spatial and
  frequency dependence ofplasma currents in a 300 mm capacitively coupled
  plasma reactor,'' {\em Plasma Sources Sci. Technol.}, vol.~15, p.~889, 2006.

\bibitem{lane_2016}
B.~Lane, C.~Campbell, I.~Sawada, and {\relax P.L.G}.~Ventzek, ``Measurement of
  spatial and temporal evolution of electromagnetic fields in a 100 {MH}z
  plasma source using b dot and double dipole probes,'' {\em J. Vac. Sci.
  Technol. A}, vol.~34, p.~031302, 2016.

\bibitem{zhao_2019}
K.~Zhao, D.-Q. Wen, Y.-X. Liu, M.~Lieberman, D.~Economou, and Y.-N. Wang,
  ``Observation of nonlinear standing waves excited by
  plasma-series-resonance-enhanced harmonics in capacitive discharges,'' {\em
  Phys. Rev. Lett.}, vol.~122, p.~185002, 2019.

\bibitem{zhao_2021}
K.~Zhao, Y.~Liu, Q.~Zhang, {\relax D.J}.~Economou, and Y.~Wang, ``Magnetic
  probe diagnostics of nonlinear standing waves and bulk ohmic electron power
  absorption in capacitive discharges,'' {\em Plasma Sources Sci. Technol.},
  vol.~23, p.~115404, 2021.

\bibitem{liu_2021a}
{\relax J.-K}.~Liu, E.~Kawamura, {\relax M.A}.~Lieberman, {\relax
  A.J}.~Lichtenberg, and {\relax Y.-N}.~Wang, ``Nonlinear harmonic excitations
  in collisional, asymmetrically-driven capacitive discharges,'' {\em Plasma
  Sources Sci. Technol.}, vol.~30, p.~045017, 2021.

\bibitem{lieberman_2015}
M.~Lieberman, A.~Lichtenberg, E.~Kawamura, and A.~Marakhtanov, ``Nonlinear
  standing wave excitation by series resonance-enhanced harmonics in low
  pressure capacitive discharges,'' {\em Plasma Sources Sci. Technol.},
  vol.~24, p.~055011, 2015.

\bibitem{annaratone_1995}
B.~Annaratone, V.~Ku, and J.~Allen, ``Identification of plasma‐sheath
  resonances in a parallel‐plate plasma reactor,'' {\em J. Appl. Phys.},
  vol.~77, p.~5455, 1995.

\bibitem{klick_1996}
M.~Klick, ``Nonlinearity of the radio‐frequency sheath,'' {\em J. Appl.
  Phys.}, vol.~79, p.~3445, 1996.

\bibitem{Ku_1998a}
{\relax V.P.T}.~Ku, {\relax B.M}.~Annaratone, and {\relax J.E}.~Allen,
  ``Plasma-sheath resonances and energy absorption phenomena in capacitively
  coupled radio frequency plasmas. part {I},'' {\em J. Appl. Phys.}, vol.~84,
  p.~6536, 1998.

\bibitem{Ku_1998b}
{\relax V.P.T}.~Ku, {\relax B.M}.~Annaratone, and {\relax J.E}.~Allen,
  ``Plasma-sheath resonances and energy absorption phenomena in capacitively
  coupled radio frequency plasmas. part {II}. the {H}erlofson paradox,'' {\em
  J. Appl. Phys.}, vol.~84, p.~6546, 1998.

\bibitem{mussenbrock_2006}
T.~Mussenbrock and R.~Brinkmann, ``Nonlinear electron resonance heating in
  capacitive radio frequency discharges,'' {\em Appl. Phys. Letters}, vol.~88,
  p.~151503, 2006.

\bibitem{czarnetzki_2006}
U.~Czarnetzki, T.~Mussenbrock, and {\relax R.P}.~Brinkmann, ``Self-excitation
  of the plasma series resonance in radio-frequency discharges: {A}n analytical
  description,'' {\em Phys. Plasmas}, vol.~13, p.~123503, 2006.

\bibitem{mussenbrock_2007}
T.~Mussenbrock and {\relax R.P}.~Brinkmann, ``Nonlinear plasma dynamics in
  capacitive radio frequency discharges,'' {\em Plasma Sources Sci. Technol.},
  vol.~16, p.~377, 2007.

\bibitem{schuengel_2015}
E.~Schüngel, S.~Brandt, I.~Korolov, A.~Derzsi, Z.~Donkó, and J.~Schulze, ``On
  the self-excitation mechanisms of plasma series resonance oscillations in
  single- and multi-frequency capacitive discharges,'' {\em Phys. Plasmas},
  vol.~22, p.~043512, 2015.

\bibitem{chabert_2007}
P.~Chabert, ``Electromagnetic effects in high-frequency capacitive discharges
  used for plasma processing,'' {\em J. Phys. D: Appl. Phys.}, vol.~40, p.~R63,
  2007.

\bibitem{lee_2008}
I.~Lee, M.~Lieberman, , and D.~Graves, ``Modeling electromagnetic effects in
  capacitive discharges,'' {\em Plasma Sources Sci. Technol.}, vol.~17,
  p.~015018, 2008.

\bibitem{kawamura_2011}
E.~Kawamura, D.~Graves, and M.~Lieberman, ``Fast 2d hybrid fluid-analytical
  simulation of inductive/capacitive discharges,'' {\em Plasma Sources Sci.
  Technol.}, vol.~20, p.~035009, 2011.

\bibitem{kawamura_2014}
E.~Kawamura, M.~Lieberman, and D.~B. Graves, ``Fast 2d fluid-analytical
  simulation of ion energy distributions and electromagnetic effects in
  multi-frequency capacitive discharges,'' {\em Plasma Sources Sci. Technol.},
  vol.~23, p.~064003, 2014.

\bibitem{rauf_2008}
S.~Rauf, K.~Bera, and K.~Collins, ``Self-consistent simulation of very high
  frequency capacitively coupled plasmas,'' {\em Plasma Sources Sci. Technol.},
  vol.~17, p.~035003, 2008.

\bibitem{bera_2009}
K.~Bera, S.~Rauf, K.~Ramaswamy, and K.~Collins, ``Effects of interelectrode gap
  on high frequency and very high frequency capacitively coupled plasmas,''
  {\em J. Vac. Sci. Technol. A}, vol.~27, p.~706, 2009.

\bibitem{yang_2010}
Y.~Yang and M.~Kushner, ``Modeling of dual frequency capacitively coupled
  plasma sources utilizing a full-wave maxwell solver: I. scaling with high
  frequency,'' {\em Plasma Sources Sci. Technol.}, vol.~19, p.~055011, 2010.

\bibitem{zhang_2010}
Y.~Zhang, S.~Zhao, A.~Bogaerts, and Y.~Wang, ``Comparison of electrostatic and
  electromagnetic simulations for very high frequency plasmas,'' {\em Phys.
  Plasmas}, vol.~17, p.~113512, 2010.

\bibitem{eremin_2013}
D.~Eremin, T.~Hemke, {\relax R.P.}.~Brinkmann, and T.~Mussenbrock,
  ``Simulations of electromagnetic effects in high-frequency capacitively
  coupled discharges using the {D}arwin approximation,'' {\em J. Phys. D: Appl.
  Phys.}, vol.~46, p.~084017, 2013.

\bibitem{wen_2017}
D.-Q. Wen, E.~Kawamura, M.~Lieberman, A.~Lichtenberg, and Y.-N. Wang, ``A
  nonlinear electromagnetics model of an asymmetrically-driven, low pressure
  capacitive discharge,'' {\em Phys. Plasmas}, vol.~24, p.~083517, 2017.

\bibitem{wen_2017a}
D.-Q. Wen, E.~Kawamura, M.~Lieberman, A.~Lichtenberg, and Y.-N. Wang,
  ``Nonlinear series resonance and standing waves in dual-frequency capacitive
  discharges,'' {\em Plasma Sources Sci. Technol.}, vol.~26, p.~015007, 2017.

\bibitem{liu_2021}
{\relax J.-K}.~Liu, {\relax Y.-R}.~Zhang, K.~Zhao, {\relax D.-Q}.~Wen, and
  {\relax Y.-N}.~Wang, ``Simulations of standing wave effect, stop band effect,
  and skin effect in large-area very high frequency symmetric capacitive
  discharges,'' {\em Plasma Sources Sci. Technol.}, vol.~23, p.~035401, 2021.

\bibitem{liu_2022}
{\relax J.-K}.~Liu, E.~Kawamura, {\relax M.A}.~Lieberman, and {\relax
  Y.-N}.~Wang, ``Nonlinear transmission line ({NTL}) model study of
  electromagnetic effects in high-frequency asymmetrically driven capacitive
  discharges,'' {\em Phys. Plasmas}, vol.~29, p.~013508, 2022.

\bibitem{rauf_2020}
S.~Rauf, ``On uniformity and non-local transport in low pressure capacitively
  coupled plasmas,'' {\em Plasma Sources Sci. Technol.}, vol.~29, p.~095019,
  2020.

\bibitem{mattei_2017}
S.~Mattei, K.~Nishid, M.~Onai, J.~Lettry, M.~Tran, and A.~Hatayama, ``A
  fully-implicit particle-in-cell {M}onte {C}arlo collision code for the
  simulation of inductively coupled plasmas,'' {\em J. Comput. Phys.},
  vol.~350, p.~891, 2017.

\bibitem{chen_2011}
G.~Chen, L.~Chac\'on, and D.~Barnes, ``An energy- and charge-conserving,
  implicit, electrostatic particle-in-cell algorithm,'' {\em J. Comput. Phys.},
  vol.~230, p.~7018, 2011.

\bibitem{markidis_2011}
S.~Markidis and G.~Lapenta, ``The energy conserving particle-in-cell method,''
  {\em J. Comput. Phys.}, vol.~230, p.~7037, 2011.

\bibitem{Eremin2022}
D.~Eremin, ``An energy- and charge-conserving electrostatic implicit
  particle-in-cell algorithm for simulations of collisional bounded plasmas,''
  {\em J. Comput. Phys}, vol.~452, p.~110934, 2022.

\bibitem{Barnes2021}
D.~Barnes and L.~Chac\'on, ``Finite spatial-grid effects in energy-conserving
  particle-in-cell algorithms,'' {\em Comput. Phys. Commun.}, vol.~258,
  p.~107560, 2021.

\bibitem{chacon_2019}
L.~Chacón and G.~Chen, ``Energy-conserving perfect-conductor boundary
  conditions for an implicit, curvilinear {D}arwin particle-in-cell
  algorithm,'' {\em J. Comput. Phys.}, vol.~391, p.~216, 2019.

\bibitem{turner_2013}
M.~Turner, A.~Derzsi, Z.~Donkó, D.~Eremin, S.~Kelly, T.~Lafleur, and
  T.~Mussenbrock, ``Simulation benchmarks for low-pressure plasmas: Capacitive
  discharges,'' {\em Phys. Plasmas}, vol.~20, p.~013507, 2013.

\bibitem{charoy_2019}
T.~Charoy, J.~Boeuf, A.~Bourdon, J.~Carlsson, P.~Chabert, B.~Cuenot, D.~Eremin,
  L.~Garrigues, K.~Hara, I.~Kaganovich, A.~Powis, A.~Smolyakov, D.~Sydorenko,
  A.~Tavant, O.~Vermorel, and W.~Villafana, ``2d axial-azimuthal
  particle-in-cell benchmark for low-temperature partially magnetized
  plasmas,'' {\em Plasma Sources Sci. Technol.}, vol.~28, p.~105010, 2019.

\bibitem{willafana_2021}
W.~Villafana, F.~Petronio, A.~Denig, M.~Jimenez, D.~Eremin, L.~Garrigues,
  F.~Taccogna, A.~Alvarez-Laguna, J.~Boeuf, A.~Bourdon, P.~Chabert, T.~Charoy,
  B.~Cuenot, K.~Hara, F.~Pechereau, A.~Smolyakov, D.~Sydorenko, A.~Tavant, and
  O.~Vermorel, ``2d radial-azimuthal particle-in-cell benchmark for $\mathbf{E}
  \times \mathbf{B}$ discharges,'' {\em Plasma Sources Sci. Technol.}, doi:
  10.1088/1361-6595/ac0a4a.

\bibitem{chacon_2013}
L.~Chac\'on, G.~Chen, and {\relax D.C}.~Barnes, ``A charge- and
  energy-conserving implicit, electrostatic particle-in-cell algorithm on
  mapped computational meshes,'' {\em J. Comput. Phys.}, vol.~233, p.~1, 2013.

\bibitem{Eremin2020}
D.~Eremin, ``Influence of the secondary electron emission on plasma uniformity
  in {VHF} low-pressure {CCP} discharges,'' in {\em 2020 IEEE International
  Conference on Plasma Science (ICOPS)}, p.~449, 2020.

\bibitem{phelps_1999}
A.~Phelps and Z.~Petrovi\'c, ``Cold-cathode discharges and breakdown in argon:
  surface and gas phase production of secondary electrons,'' {\em Plasma
  Sources Sci. Technol.}, vol.~8, p.~R21, 1999.

\bibitem{phelps_1994}
A.~Phelps, ``The application of scattering cross sections to ion flux models in
  discharge sheaths,'' {\em J. Appl. Phys.}, vol.~76, p.~747, 1994.

\bibitem{mertmann_2011}
P.~Mertmann, D.~Eremin, T.~Mussenbrock, R.~Brinkmann, and P.~Awakowicz,
  ``Fine-sorting one-dimensional particle-in-cell algorithm with monte-carlo
  collisions on a graphics processing unit,'' {\em Comp. Phys. Comm.},
  vol.~182, no.~10, p.~2161, 2011.

\bibitem{Marsaglia_2003}
G.~Marsaglia, ``Xorshift rngs,'' {\em Journal of Statistical Software}, vol.~8,
  no.~14, p.~1, 2003.

\bibitem{yee_1966}
K.~Yee, ``Numerical solution of initial boundary value problems involving
  {M}axwell’s equations in isotropic media,'' {\em IEEE Trans. Antennas
  Propag.}, vol.~14, no.~3, p.~302, 1966.

\bibitem{eremin_2021b}
D.~Eremin, B.~Berger, D.~Engel, J.~Kallähn, K.~Köhn, D.~Krüger, L.~Xu,
  M.~Oberberg, C.~Wölfel, J.~Lunze, P.~Awakowicz, J.~Schulze, and
  R.~Brinkmann, ``Electron dynamics in planar radio frequency magnetron
  plasmas: {II}. {H}eating and energization mechanisms studied via a 2d3v
  particle-in-cell/{M}onte {C}arlo code,'' {\em Manuscript submitted for
  publication}.

\bibitem{delzanno_2013b}
{\relax G.L}.~Delzanno and E.~Camporeale, ``On particle movers in cylindrical
  geometry for particle-in-cell simulations,'' {\em J. Comput. Phys.},
  vol.~253, p.~259, 2013.

\bibitem{wang_1999}
J.~Wang, D.~Kondrashov, P.~Liewer, and S.~Karmesin, ``Three-dimensional
  deformable-grid electromagnetic particle-in-cell for parallel computers,''
  {\em Journal of Plasma Physics}, vol.~61, no.~3, p.~367, 1999.

\bibitem{delzanno_2013}
G.~Delzanno, E.~Camporeale, J.~Moulton, and M.~Thomsen, ``{CPIC}: A curvilinear
  particle-in-cell code for plasma–material interaction studies,'' {\em IEEE
  Transactions on Plasma Science}, vol.~41, no.~12, p.~3577, 2013.

\bibitem{mur_1981}
G.~Mur, ``Absorbing boundary conditions for the finite-difference approximation
  of the time-domain electromagnetic-field equations,'' {\em IEEE Trans.
  Electromagn. Compat.}, vol.~EMC-23, p.~377, 1981.

\bibitem{alvarez_2007}
R.~Alvarez and {\relax L.L}.~Alves, ``Two-dimensional electromagnetic model of
  a microwave plasma reactor operated by an axial injection torch,'' {\em J.
  Appl. Phys.}, vol.~101, p.~103303, 2007.

\bibitem{jimenez-diaz_2011}
M.~Jimenez-Diaz, J.~van Dijk, and {\relax J.J.A.M}.~van~der Mullen, ``Effect of
  remote field electromagnetic boundary conditions on microwave-induced plasma
  torches,'' {\em J. Phys. D: Appl. Phys.}, vol.~44, p.~165203, 2011.

\bibitem{rahimi_2014}
S.~Rahimi, M.~Jimenez-Diaz, S.~Hübner, {\relax E.H}.~Kemaneci, {\relax
  J.J.A.M}.~van~der Mullen, and J.~van Dijk, ``A two-dimensional modelling
  study of a coaxial plasma waveguide,'' {\em J. Phys. D: Appl. Phys.},
  vol.~47, p.~125204, 2014.

\bibitem{chen_2014}
G.~Chen and L.~Chacón, ``An energy- and charge-conserving, nonlinearly
  implicit, electromagnetic 1d-3v {V}lasov–{D}arwin particle-in-cell
  algorithm,'' {\em Comput. Phys. Commun.}, vol.~185, no.~10, p.~2391, 2014.

\bibitem{chen_2015}
G.~Chen and L.~Chacón, ``A multi-dimensional, energy- and charge-conserving,
  nonlinearly implicit, electromagnetic vlasov–darwin particle-in-cell
  algorithm,'' {\em Comput. Phys. Commun.}, vol.~197, p.~73, 2015.

\bibitem{chacon_2016}
L.~Chacón and G.~Chen, ``A curvilinear, fully implicit, conservative
  electromagnetic pic algorithm in multiple dimensions,'' {\em J. Comput.
  Phys.}, vol.~316, p.~578, 2016.

\bibitem{yamazawa_2015}
Y.~Yamazawa, ``Electrode impedance effect in dual-frequency capacitively
  coupled plasma,'' {\em Plasma Sources Sci. Technol.}, vol.~24, p.~034015,
  2015.

\bibitem{schmidt_2018}
F.~Schmidt, J.~Schulze, E.~Johnson, {\relax J.-P}.~Booth, D.~Keil, {\relax
  D.M}.~French, J.~Trieschmann, and T.~Mussenbrock, ``Multi frequency matching
  for voltage waveform tailoring,'' {\em Plasma Sources Sci. Technol.},
  vol.~27, p.~095012, 2018.

\bibitem{schmidt_2019}
F.~Schmidt, J.~Trieschmann, T.~Gergs, and T.~Mussenbrock, ``A generic method
  for equipping arbitrary rf discharge simulation frameworks with external
  lumped element circuits,'' {\em J. Appl. Phys.}, vol.~125, p.~173106, 2019.

\bibitem{verboncoeur_1993}
J.~Verboncoeur, M.~Alves, V.~Vahedi, and C.~Birdsall, ``Simultaneous potential
  and circuit solution for 1d bounded plasma particle simulation codes,'' {\em
  J. Comput. Phys.}, vol.~104, p.~321, 1993.

\bibitem{vahedi_1997}
V.~Vahedi and G.~DiPeso, ``Simultaneous potential and circuit solution for
  two-dimensional bounded plasma simulation codes,'' {\em J. Comput. Phys.},
  vol.~131, p.~141, 1997.

\bibitem{eremin_2016b}
D.~Eremin, T.~Hemke, and T.~Mussenbrock, ``A new hybrid scheme for simulations
  of highly collisional rf-driven plasmas,'' {\em Plasma Sources Sci.
  Technol.}, vol.~25, p.~015009, 2016.

\bibitem{brackbill_2015}
{\relax J.U}.~Brackbill, ``On energy and momentum conservation in
  particle-in-cell plasma simulation,'' {\em J. Comput. Phys.}, vol.~317,
  p.~405, 2016.

\bibitem{taitano_2013}
W.~Taitano, D.~Knoll, L.~Chac\'on, and G.~Chen, ``Development of a consistent
  and stable fully implicit moment method for {V}lasov–{A}mp\`ere particle in
  cell ({PIC}) system,'' {\em SIAM J. Sci. Comput.}, vol.~35, no.~5, p.~S126,
  2013.

\bibitem{Lieberman_2022}
{\relax M.A}.~Lieberman, E.~Kawamura, and P.~Chabert, ``Standing wave
  instability in large area capacitive discharges operated within or near the
  gamma mode,'' {\em Plasma Sources Sci. Technol.}, vol.~31, p.~114007, 2022.

\bibitem{wilczek_2016}
S.~Wilczek, J.~Trieschmann, D.~Eremin, R.~Brinkmann, J.~Schulze, E.~Schuengel,
  A.~Derzsi, I.~Korolov, P.~Hartmann, Z.~Donkó, and T.~Mussenbrock, ``Kinetic
  interpretation of resonance phenomena in low pressure capacitively coupled
  radio frequency plasmas,'' {\em Phys. Plasmas}, vol.~23, p.~063514, 2016.

\bibitem{wilczek_2018}
S.~Wilczek, J.~Trieschmann, J.~Schulze, Z.~Donkó, {\relax R.P}.~Brinkmann, and
  T.~Mussenbrock, ``Disparity between current and voltage driven capacitively
  coupled radio frequency discharges,'' {\em Plasma Sources Sci. Technol.},
  vol.~27, p.~125010, 2018.

\bibitem{perret_2005}
A.~Perret, P.~Chabert, J.~Jolly, and {\relax J.-P}.~Booth, ``Ion energy
  uniformity in high-frequency capacitive discharges,'' {\em Appl. Phys.
  Lett.}, vol.~86, p.~021501, 2005.

\bibitem{howling_2005a}
{\relax A.A}.~Howling, L.~Sansonnens, H.~Schmidt, and {\relax Ch}.~Hollenstein,
  ``Comment on ``{I}on energy uniformity in high-frequency capacitive
  discharges'' {[Appl. Phys. Lett. 86, 021501 (2005)]},'' {\em Appl. Phys.
  Lett.}, vol.~87, p.~076101, 2005.

\bibitem{anders_2014}
A.~Anders, ``Localized heating of electrons in ionization zones: Going beyond
  the {P}enning-{T}hornton paradigm in magnetron sputtering,'' {\em Appl. Phys.
  Lett.}, vol.~105, p.~244104, 2014.

\bibitem{eremin_2021a}
D.~Eremin, D.~Engel, D.~Krüger, S.~Wilczek, B.~Berger, M.~Oberberg,
  C.~Wölfel, J.~Lunze, P.~Awakowicz, J.~Schulze, and R.~Brinkmann, ``Electron
  dynamics in planar radio frequency magnetron plasma: {I. T}he mechanism of
  {H}all heating,'' {\em Manuscript submitted for publication}.

\bibitem{lieberman_2008}
{\relax M.A}.~Lieberman, {\relax A.J}.~Lichtenberg, E.~Kawamura,
  T.~Mussenbrock, and {\relax R.P}.~Brinkmann, ``The effects of nonlinear
  series resonance on {O}hmic and stochastic heating in capacitive
  discharges,'' {\em Phys. Plasmas}, vol.~15, p.~063505, 2008.

\bibitem{popov_1985}
{\relax O.A}.~Popov and {\relax V.A}.~Godyak, ``Power dissipated in
  low‐pressure radio‐frequency discharge plasmas,'' {\em J. Appl. Phys.},
  vol.~57, p.~53, 1985.

\bibitem{schulze_2015}
J.~Schulze, Z.~Donkó, A.~Derzsi, I.~Korolov, and E.~Schuengel, ``The effect of
  ambipolar electric fields on the electron heating in capacitive rf plasmas,''
  {\em Plasma Sources Sci. Technol.}, vol.~24, p.~015019, 2015.

\bibitem{kaganovich_1996}
{\relax I.D}.~Kaganovich, {\relax V.I}.~Kolobov, and {\relax L.D}.~Tsendin,
  ``Stochastic electron heating in bounded radio‐frequency plasmas,'' {\em
  Appl. Phys. Lett.}, vol.~69, p.~3818, 1996.

\bibitem{lafleur_2014}
T.~Lafleur, P.~Chabert, and {\relax J.P}.~Booth, ``Electron heating in
  capacitively coupled plasmas revisited,'' {\em Plasma Sources Sci. Technol.},
  vol.~23, p.~035010, 2014.

\bibitem{lafleur_2015}
T.~Lafleur and P.~Chabert, ``Is collisionless heating in capacitively coupled
  plasmas really collisionless?,'' {\em Plasma Sources Sci. Technol.}, vol.~24,
  p.~044002, 2015.

\bibitem{berger_2021}
B.~Berger, D.~Eremin, M.~Oberberg, D.~Engel, C.~W\"{o}lfel, Q.-Z. Zhang,
  P.~Awakowicz, J.~Lunze, R.~Brinkmann, and J.~Schulze, ``Electron dynamics in
  planar radio frequency magnetron plasma: {III. P}ower absorption dynamics in
  capacitively coupled radio-frequency magnetrons with a conducting target,''
  {\em Manuscript submitted for publication}.

\bibitem{Schroeder_1993}
J.~Schroeder, ``Signal processing via {F}ourier-{B}essel series expansion,''
  {\em Digital Signal Processing}, vol.~3, no.~2, p.~112, 1993.

\end{thebibliography}
